\documentclass[aps,floats,showpacs,nofootinbib,floatfix,10pt]{revtex4}%
\usepackage{makeidx}
\usepackage{eurosym}
\usepackage{amsfonts}
\usepackage{amsmath}
\usepackage{graphicx}
\usepackage{amssymb}%
{\catcode`\|=\active \gdef\Braket#1{\left<\mathcode`\|"8000\let|\bravert
{#1}\right>}}
\def\bravert{\egroup\,\vrule\,\bgroup}

\begin{document}
\title{Non local lagrangians: the pion.}
\author{S. Noguera}
\email{Santiago.Noguera@uv.es}
\affiliation{Departamento de Fisica Teorica and Instituto de F\'{\i}sica Corpuscular,
Universidad de Valencia-CSIC, E-46100 Burjassot (Valencia), Spain.}
\date{\today }

\begin{abstract}

\end{abstract}
\begin{abstract}
We define a family of non local and chirally symmetric low energy lagrangians
motivated by theoretical studies on Quantum Chromodynamics. These models lead
to quark propagators with non trivial momentum dependencies. We define the
formalism for two body bound states and apply it to the pion. We study the
coupling of the photon and W bosons with special attention to the
implementation of local gauge invariance. We calculate the pion decay constant
recovering the Goldberger-Treiman and the Gell-Mann-Oakes-Renner relations. We
recover a form of the axial current consistent with PCAC. Finally we study the
pion form factor and we construct the operators involved in its parton distribution.

\end{abstract}

\pacs{24.10.Jv, 11.10.St, 13.40.Gp, 13.60.Fz}
\maketitle

\section{Introduction.}

The strong interaction among hadrons is supposed to be described by Quantum
Chromodynamics (QCD) \cite{fgl} which is a field theory defined in terms of
quark and gluon fields. While the asymptotic behavior of QCD is well
understood and its proponents worthy of the highest recognition \cite{nobel},
the low energy behavior is still a subject of much scientific endeavor. Low
energy physics seems to be ultimately governed by flavor dynamics. Confinement
\cite{wilson}, the property of QCD which describes how the dynamics based on
color in the lagrangian transforms into a dynamics based on flavor for the
physical states, and why these cannot exist with color charge, is still a
subject of research and debate. This complex low energy behavior is described
conventionally in terms of approximations to the theory, i.e., lattice QCD
\cite{lattice}, non relativistic QCD \cite{nrqcd}, $1/N_{c}$ -expansion
\cite{hooft} or effective theories, i.e. Chiral Perturbation Theory \cite{gl},
Heavy Quark Effective Theory \cite{hqet}, etc. Models turn out to be extremely
useful in some instances when the other approaches are too complex, i.e,
non-relativistic quark models \cite{ik,mg}, bag models \cite{mit,vento},
Nambu-Jona Lasinio (NJL) model \cite{njl}, chiral-soliton models
\cite{skyrme,vento1,diakonov}, etc. Another method to study non perturbative
physics in a lagrangian theory like QCD is to solve the Dyson-Schwinger
equations. The application of this formalism to QCD becomes an enormously
difficult task but progress, in understanding the theory from this approach,
has been achieved~\cite{RobertsWilliams94-97, AlkoferSmekal01, MarisRoberts03}%
. The global color model \cite{Tandy97}, the\ extended non-local NJL model
\cite{Birse95, Birse98, Birse04,Osipov95,Osipov06}, and models using separable
interactions \cite{GrossBuckIto91} has been introduced as model realizations
of QCD in a field theory formalism.

One major problem is to understand the pion, because it is the system which
contains all the ingredient of QCD: asymptotic freedom, confinement and
spontaneously broken chiral symmetry. Chiral symmetry governs the static
properties of the theory, like the quark condensate, the mass and decay
constant of the pion. The dynamics fixes the internal structure of the pion,
which is accessible through the pion electromagnetic form factor.

The pion form factor has been a subject of many studies. In relativistic
quantum mechanics the pion form factor has a long history of debate
\cite{IsgurLlewellyn, ChungKoesterPolitzou, CardarelliPaceSalmeSimula, ChoiJi,
MeloNausFrederico}. One of the main problems is the choice of one of the
various Dirac forms \cite{DesplanquesNogueraTheussl,AmgharDesplanquesTheussl}.
One of the most interesting conclusions is that results of a calculation
depend not only on the Dirac form chosen but also on the frame chosen
\cite{TheusslAmgharDesplanquesNog03,Desplanques}. The reason for this is that
the truncation needed in relativistic quantum mechanics to calculate form
factors breaks Poincar\'{e} invariance.

A way to avoid this problem is to work in a field theory formalism. The pion
form factor has also been studied within Dyson-Schwinger Equations schemes by
several authors \cite{Roberts96, BurdenRobertsThomson96, MarisTandi00}. The
starting point is in most treatments the pion Bether-Salpeter amplitude
calculated in the rainbow approximation. The pion form factor is calculated
using the so called impulse approximation which considers only the triangle
diagram, and the use of a dressed vertex for the photon. For the latter the
Ball-Chiu \cite{BallChiu80} expression for the vertex , or modified versions
of it, have been used.

The pion form factor has been studied also for the non-local NJL model
\cite{Birse98}. In this case the coupling of the photon is obtained
by\ restoration of the gauge symmetry \cite{Birse95, Bos91, Osipov06}.

Our aim is to construct a model for hadron structure and hadronic interactions
developing a formalism which preserves the fundamental symmetries of the
theory\ (chiral, Poincar\'{e} and local electromagnetic gauge invariances) and
which incorporates information coming from fundamental studies of Quantum
Chromodynamics. For this purpose we want to define a formalism which contains
the physical intuition of model calculations and the lagrangian formalism of
the effective theories. To achieve this we have found that the best suited
scheme is to describe the physics by mean of a phenomenological chirally
invariant non local lagrangian.

Working in a lagrangian theory, the two main ingredients in a non perturbative
analysis involving the pion are: i) the quark propagator, obeying the Dyson
equation; ii) the description of the pion as a bound state of a Bethe-Salpeter
equation (BSE). Due to chiral symmetry the kernels of these two equations are
not independent \cite{DelbourgoScadron79}. Solving the Dyson equation for our
lagrangian leads to momentum dependencies in the quark propagators through its
mass and its wave function renormalization. In our scheme the gluons have been
integrated out and we have only flavor interaction between quarks. Confinement
is imposed by the structure of the quark propagator and by limiting the Fock
space to color singlet states. The pion is obtained in a consistent way
solving the BSE, and the Goldstone character of the pion is recovered.

Our model can be seen as an extension of the non-local NJL model
\cite{Birse95, Birse98, Osipov95, Osipov06}, but with a particular philosophy.
We consider the description of the quark propagator as the main ingredient.
This is because the quark propagator is the first information that can be
obtained from fundamental studies, as lattice QCD. Our lagrangian is the
minimal extension which allows to incorporate the full momentum dependence of
the quark propagator, through its mass and wave function renormalization. From
this lagrangian we can explore what are the implications for other observables
originated by changes in the quark propagator.

Our formalism implements the coupling of the photon in a gauge invariant
manner\textbf{\textbf{ }}\cite{Birse95, Bos91, Osipov06}. This allows to study
the electromagnetic properties of the pion, which depend strongly on the
quark-photon vertex. Usually this vertex is calculated by using the
Ward-Takahashi identity \cite{BallChiu80}. This method fixes the longitudinal
part of the vertex leaving the transverse part unconstrained. We show that
this procedure does not guarantee local gauge invariance, while ours does. As
an application we study the pion form factor, showing that the conventional
impulse approximation, in which the form factor is calculated using the
triangle diagram, is not consistent with local gauge symmetry.

In models based on field theory formalism, the construction of the axial
current and the definition of the pion decay constant need particular
attention. In ref \cite{MarisRobertsTandy98} a first expression for the axial
current is given. In ref \cite{Birse95, Osipov06} additional contributions to
the pion decay constant are included. In this paper we implement the coupling
of quarks to the $W^{\mu}$\textbf{ }bosons in a gauge invariant manner
following a procedure similar to the one used for photons. Then, we analyze
the axial current and the pion decay constant.

Our formalism is very effective for building operators describing observables
in a consistent way. As an application we have studied the operators involved
in the parton distribution of the pion.

This paper is organized as follows. In section \ref{SecLagrangian} we define
our lagrangian, we discuss the quark propagator, and we fix the parameters in
order to describe the adequate quark propagator obtained by more fundamental
studies based on QCD. In section \ref{SecPionMass} we describe the pion state.
In section \ref{SecQuarkPhotonVertex} we study the quark-photon vertex and
recover the Ball-Chiu ansatz but with additional contributions. In section
\ref{SecAxialCurrent_fpi} we study the axial current and the pion decay
constant. In section \ref{SecPionFF} we apply the model to the study of the
pion form factor. We show that a four quark-one photon vertex appears in a
natural way. In section \ref{SecPionPD} we obtain the contribution of this new
term to the parton distribution operator. The last section contains the
conclusions of our investigation.

\section{A phenomenological non local Lagrangian for hadron structure.}

\label{SecLagrangian}
\renewcommand{\theequation}{\thesection.\arabic{equation}} \setcounter{equation}{0}

Let us build a model which produces a non trivial momentum dependence in the
quark propagator and preserves all the required symmetries: Poincar\'{e} and
chiral symmetry. This momentum dependence will arise from a lagrangian
description and manifests itself as a quark momentum dependent mass and a
quark momentum dependent wave function renormalization. Let us define the non
local currents as
\begin{equation}
J_{\mathcal{O}}\left(  x\right)  =\int d^{4}y\,G\left(  y\right)  \bar{\psi
}\left(  x+\frac{1}{2}y\right)  \mathcal{O}\psi\left(  x-\frac{1}{2}y\right)
\ , \label{02.01NonLocalCurrents}%
\end{equation}
where the operator $\mathcal{O}$ is such that
\begin{equation}
\gamma^{0}\mathcal{O}^{\dagger}\gamma^{0}=\mathcal{O\ }. \label{02.02}%
\end{equation}
With these definitions the hermiticity of the currents $\left(  J_{\mathcal{O}
}^{\dagger}\left(  x\right)  =J_{\mathcal{O}}\left(  x\right)  \right)  $
implies that $G^{\dagger}\left(  -y\right)  =G\left(  y\right)  .$ A local
current corresponds to $G\left(  y\right)  =\delta^{4}\left(  y\right)  $ and
thus the natural normalization for the functions $G\left(  y\right)  $ is
\begin{equation}
\int d^{4}y\,G\left(  y\right)  =1. \label{02.03}%
\end{equation}

We build a lagrangian in terms of non local currents, preserving $U_{L}\left(
2\right)  \otimes U_{R}\left(  2\right)  $ chiral symmetry, and producing the
desired momentum dependences as
\begin{equation}
\mathcal{L}\left(  x\right)  =\bar{\psi}\left(  x\right)  \left(
i\ \rlap{$/$}\partial-m_{0}\right)  \psi\left(  x\right)  +g_{0}\left[
J_{S}^{\dagger}\left(  x\right)  J_{S}\left(  x\right)  +\vec{J}%
_{5}^{\;\dagger}\left(  x\right)  \vec{J}_{5}\left(  x\right)  \right]
+g_{p}J_{p}^{\dagger}\left(  x\right)  J_{p}\left(  x\right)
\ ,\label{02.04Lagrangian}%
\end{equation}
where the currents are defined by
\begin{align}
J_{S}\left(  x\right)   &  =\int d^{4}y\,G_{0}\left(  y\right)  \bar{\psi
}\left(  x+\frac{1}{2}y\right)  \psi\left(  x-\frac{1}{2}y\right)
\ ,\label{02.05ScalarCurrent}\\
\vec{J}_{5}\left(  x\right)   &  =\int d^{4}y\,G_{0}\left(  y\right)
\bar{\psi}\left(  x+\frac{1}{2}y\right)  i\vec{\tau}\gamma_{5}\psi\left(
x-\frac{1}{2}y\right)  \ ,\label{02.06PseudoscalarCurrent}\\
J_{p}\left(  x\right)   &  =\int d^{4}y\,G_{p}\left(  y\right)  \bar{\psi
}\left(  x+\frac{1}{2}y\right)  \frac{1}{2}i\overleftrightarrow
{\rlap{$/$}\partial}\psi\left(  x-\frac{1}{2}y\right)
\ ,\label{02.07MomentumCurrent}%
\end{align}
where $u\overleftrightarrow{\partial}v=u\left(  \partial v\right)  -\left(
\partial u\right)  v.$ The transformation properties of the non local currents
are the same as those of the local ones. The first and second currents require
the same $G_{0}\left(  y\right)  $ to guarantee chiral invariance. The third
current is self-invariant under chiral transformations. The scalar current,
$J_{S},$ generates a momentum dependent mass, and the last current, the
"momentum" current, $J_{p},$ is responsible for the momentum dependence of the
wave function renormalization. The pseudo-scalar current, $\vec{J}_{5},$
generates the pion pole. From now on, just for simplicity, we assume that all
the $G\left(  y\right)  $ functions are real.\footnote{In a previous paper
another derivative coupling, $\left(  \bar{\psi}\left(  x\right)  \frac{1}%
{2}i\overleftrightarrow{\partial^{\mu}}\psi\left(  x\right)  \right)  ^{2},$
was introduced \cite{Osipov95}. This term is built with vector currents and
therefore, it produces different effects than our term $J_{p}^{2}$. In
particular, it was used to reproduce vector meson dominance.}

The interaction vertex obtained from the Lagrangian (\ref{02.04Lagrangian})
automatically includes vertex form factors. Let us define
\begin{equation}
G\left(  p\right)  =\int d^{4}y\,e^{iyp}G\left(  y\right)  \ \ , \label{02.08}%
\end{equation}
with the normalization condition
\begin{equation}
G\left(  p=0\right)  =1. \label{02.09}%
\end{equation}

\bigskip

\begin{figure}[ptb]
\begin{center}
\includegraphics[height=0.7187in,width=3.9967in]
{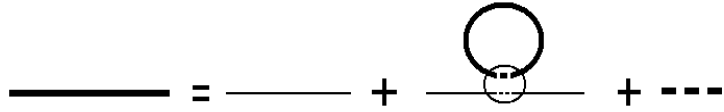}
\end{center}
\caption{Diagrammatic representation of the quark propagator. The non local
four quark vertex is represented by a circle with structure.}%
\label{FigquarkProp}%
\end{figure}The full quark propagator is obtained from de Dyson equation,
represented in Fig. \ref{FigquarkProp},%
\begin{equation}
S\left(  p\right)  =\frac{1}{\rlap{$/$}p-m_{0}+\Sigma\left(  p\right)
+i\epsilon}\ \ ,\label{03.04}%
\end{equation}
with
\begin{equation}
\Sigma\left(  p\right)  =-\alpha_{0}\ G_{0}\left(  p\right)
-\rlap{$/$}p\ \alpha_{p}\ G_{p}\left(  p\right)  \ \ ,\label{03.05}%
\end{equation}
where the first term arises from the scalar current and the second from the
momentum current. The constants $\alpha_{0}$ and $\alpha_{p}$ are directly
related to the couplings $g_{0}$ and $g_{p}$,
\begin{align}
\alpha_{0} &  =2\,g_{0}\int\frac{d^{4}p}{\left(  2\pi\right)  ^{4}}%
G_{0}\left(  p\right)  \mathbb{T}\text{r}\left(  iS\left(  p\right)  \right)
=i\,8\,N_{c}N_{f}g_{0}\int\frac{d^{4}p}{\left(  2\pi\right)  ^{4}}G_{0}\left(
p\right)  \frac{Z\left(  p\right)  m\left(  p\right)  }{p^{2}-m^{2}\left(
p\right)  +i\epsilon}\ \ ,\label{03.08}\\
\alpha_{p} &  =2\,g_{p}\int\frac{d^{4}p}{\left(  2\pi\right)  ^{4}}%
G_{p}\left(  p\right)  \mathbb{T}\text{r}\left(  iS\left(  p\right)
\rlap{$/$}p\right)  =i\,8\,N_{c}N_{f}g_{p}\int\frac{d^{4}p}{\left(
2\pi\right)  ^{4}}G_{p}\left(  p\right)  \frac{p^{2}Z\left(  p\right)  }%
{p^{2}-m^{2}\left(  p\right)  +i\epsilon}\ \ ,\label{03.09}%
\end{align}
where $\mathbb{T}$r represents the trace in Dirac, color and flavor indices.
For simplicity we work from now in the large $N_{c}$ limit. This is equivalent
to the Hartree approximation\ which implies that only direct terms are taken
into account.

We can rewrite the momentum dependence of the quark propagator in a more
standard way through a momentum dependence in the quark mass and in the quark
wave function renormalization,%
\begin{equation}
S\left(  p\right)  =Z\left(  p\right)  \frac{\rlap{$/$}p+m\left(  p\right)
}{p^{2}-m^{2}\left(  p\right)  +i\epsilon}~~. \label{03.01QuarkPropagator}%
\end{equation}
with%
\begin{align}
m\left(  p\right)   &  =\frac{m_{0}+\alpha_{0}G_{0}\left(  p\right)
}{1-\alpha_{p}G_{p}\left(  p\right)  }\ \ ,\label{03.07Mass}\\
Z\left(  p\right)   &  =\frac{1}{1-\alpha_{p}G_{p}\left(  p\right)  }\ \ .
\label{03.06WaveNormalization}%
\end{align}
These relations between $\left(  G_{0}\left(  p\right)  \text{ and }%
G_{p}\left(  p\right)  \right)  $ and $\left(  m\left(  p\right)  \text{ and
}Z\left(  p\right)  \right)  $ assures the self consistency of the solution of
the Dyson equation.

Eqs. (\ref{03.07Mass}) and (\ref{03.06WaveNormalization}) show that the scalar
current can give a mass to the quark even if the lagrangian contains no mass
term, $m_{0}=0.$ This phenomenon is the spontaneous symmetry breaking
mechanism which is similar to that taking place in the Nambu-Jona Lasinio
model \cite{njl}. On the other hand, the momentum current gives rise to a
momentum dependent wave function normalization. However, although it
contributes to the mass, it is not able by itself to break spontaneously
chiral symmetry.

\bigskip

We shall be guided by fundamental studies of QCD and lattice parametrizations
for building models for $G_{0}\left(  p\right)  $\textbf{ and }$G_{p}\left(
p\right)  $. The natural way to proceed is to use the information coming from
these studies to write ans\"{a}tze for $m\left(  p\right)  $\ and $Z\left(
p\right)  $. Then, transposing Eqs.(\ref{03.07Mass}) and
(\ref{03.06WaveNormalization}) we obtain $G_{0}\left(  p\right)  $\ and
$G_{p}\left(  p\right)  .$\ The values for $\alpha_{0}$\ and $\alpha_{p}$\ are
determined from the normalization condition equation (\ref{02.09})
\begin{align}
\alpha_{0}  &  =\frac{m\left(  0\right)  }{Z\left(  0\right)  }-m_{0}%
\ \ ,\label{03.070Mass0}\\
\alpha_{p}  &  =1-\frac{1}{Z\left(  0\right)  }\ \ ,
\label{03.060WaveNormalization0}%
\end{align}
and, from Eqs. (\ref{03.08}) and (\ref{03.09}), we obtain the values for
$g_{0}$\ and $g_{p}.$

These studies are performed in Euclidean space and therefore we will perform
our calculations in this space. We use $p_{E}$\ to represent the momentum in
Euclidean space.

Here $G_{0}\left(  p_{E}\right)  ,$ $G_{p}\left(  p_{E}\right)  ,$ $Z\left(
p_{E}\right)  $ and $m\left(  p_{E}\right)  $ are functions of $p_{E}^{2}.$ We
impose that for $p_{E}^{2}\rightarrow\infty,$ the mass goes to the current
mass and the wave function renormalization to 1,
\begin{align}
&  m\left(  p_{E}\right)  \underset{p_{E}^{2}\rightarrow\infty}%
{\longrightarrow}m_{0}~,\label{03.18}\\
&  Z\left(  p_{E}\right)  \underset{p_{E}^{2}\rightarrow\infty}%
{\longrightarrow}1~. \label{03.19}%
\end{align}
Assuming that the integrals in Eqs. (\ref{03.08}) and (\ref{03.09}) are
convergent, and looking at the behavior of the integrands for large values of
$p_{E}^{2}$ we obtain that
\begin{align}
G_{0}\left(  p_{E}\right)  \underset{p_{E}^{2}\rightarrow\infty}%
{\longrightarrow}p_{E}^{-\alpha}~~\text{with~~}\alpha &  >2+\varepsilon
\label{03.20}\\
G_{p}\left(  p_{E}\right)  \underset{p_{E}^{2}\rightarrow\infty}%
{\longrightarrow}p_{E}^{-\alpha}~~\text{with~~}\alpha &  >4+\varepsilon~.
\label{03.21}%
\end{align}

Let us define $G_{0}\left(  p_{E}\right)  $\ and $G_{p}\left(  p_{E}\right)
$\ or alternatively $m\left(  p_{E}\right)  $\ and $Z\left(  p_{E}\right)
.$\ Much research has been carried out in the study of their functional
shapes. We extract from these studies two well known scenarios based on
different philosophies but equally consistent.

The first scenario, which we will call S1, is based on the work of Dyakonov
and Petrov \cite{DyakonovPetrov86}. They provide us with the momentum
dependence of the quark mass term coming from an instanton model. They assume
$Z\left(  p_{E}\right)  =1$ and work in the chiral limit ($m_{0}=0$). Their
results are well described by the expression
\begin{equation}
m\left(  p_{E}\right)  =m_{0}+\alpha_{m}\left(  \frac{\Lambda_{m}^{2}}%
{\Lambda_{m}^{2}+p_{E}^{2}}\right)  ^{3/2}\ \ , \label{03.22DyakonovPetrov0}%
\end{equation}
with $\Lambda_{m}=0.767\operatorname{GeV}$ and $\alpha_{m}%
=0.343\operatorname{GeV}.$

The second scenario, which we call S2, corresponds to an alternative mass
function obtained from\ lattice calculations as proposed by Bowman et al.
\cite{Bowman02, Bowman03},
\begin{equation}
m\left(  p_{E}\right)  =m_{0}+\alpha_{m}\frac{\Lambda_{m}^{3}}{\Lambda_{m}%
^{3}+\left(  p_{E}^{2}\right)  ^{1.5}}\ \ ,\label{03.23mBowman}%
\end{equation}
with $\Lambda_{m}=0.719\operatorname{GeV}$ and $\alpha_{m}%
=0.302\operatorname{GeV}.$ In their lattice analysis the authors also look for
the wave function renormalization constant. Their values are reasonably
reproduced by
\begin{equation}
Z\left(  p_{E}\right)  =1+\alpha_{z}\left(  \frac{\Lambda_{z}^{2}}{\Lambda
_{z}^{2}+p_{E}^{2}}\right)  ^{5/2}\ \ ,\label{03.24ZBowman}%
\end{equation}
with $\alpha_{z}=-0.5$ and $\Lambda_{z}=1.183\operatorname{GeV}.$

In table \ref{Table 1} we show the values of some observables for the
different scenarios. Among them, the quark condensate is defined by
\begin{equation}
\left\langle \bar{q}q\right\rangle =-i\,4\,N_{c}\int\frac{d^{4}p}{\left(
2\pi\right)  ^{4}}\left(  \frac{Z\left(  p\right)  m\left(  p\right)  }%
{p^{2}-m^{2}\left(  p\right)  +i\epsilon}-\frac{m_{0}}{p^{2}-m_{0}%
^{2}+i\epsilon}\right)  \ \ . \label{03.11}%
\end{equation}
We stress that these values are obtained without any free parameter and
therefore they are model predictions.\begin{table}[ptb]
\centering%
\begin{tabular}
[c]{|c|c|c|c|c|c|}\hline
$\ $Case$\ $ & $\left\langle \bar{q}q\right\rangle ^{1/3}\left(
\operatorname{MeV}\right)  $ & $m_{0}\left(  \operatorname{MeV}\right)  $ &
$m_{\pi}\left(  \operatorname{MeV}\right)  $ & $f_{\pi}(\operatorname{MeV})$ &
$<r^{2}>(\operatorname{fm}^{2})$\\\hline
S1 & $-303.$ & 2.3 & $137.$ & 82. & $0.41~~\left(  0.41\right)  $\\\hline
S2 & $-285.$ & 3.0 & $139.$ & 81. & $0.36~~\left(  0.40\right)  $\\\hline
Exp. & $-250.\sim-300.$ & $1.5\sim8.$ & $135.\sim140.$ & 92. & $0.44$\\\hline
\end{tabular}
\caption{Results for $<qq>^{1/3}$ , $m_{0}$, the corresponding pion mass, the
pion decay constant and the mean square radius, $<r^{2}>$, for the full
vertices given by Eq. (\ref{05.12DressedQuarkPhoton}) and, between brakets,
the Ball-Chiu ansatz for the two scenarios described in the main text.}%
\label{Table 1}%
\end{table}

\section{The Pion mass.}

\label{SecPionMass}
\begin{figure}[ptb]
\begin{center}
\includegraphics[height=3.2895cm,width=13.7226cm]
{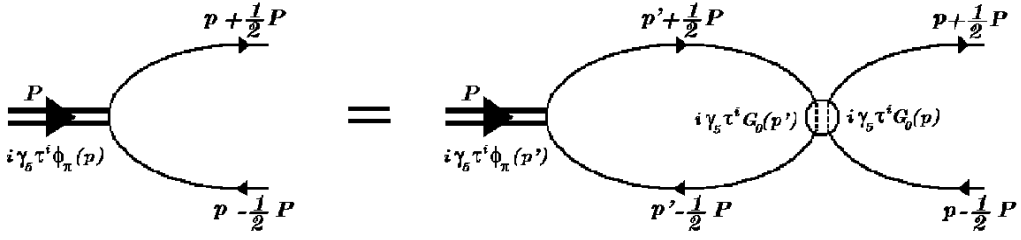}
\end{center}
\caption{Diagramatic representation of the Bethe Salpeter equation.}%
\label{FigBetheSalpeter}%
\end{figure}

\setcounter{equation}{0}In our formalism the Bethe-Salpeter
amplitude in the two body pion channel is defined as
\begin{equation}
\chi^{i}\left(  p,P\right)  =i\,S\left(  p+\frac{1}{2}P\right)  \,i\,\gamma
_{5}\,\tau^{i}\,\phi_{\pi}\left(  p\right)  \,i\,S\left(  p-\frac{1}%
{2}P\right)  \ \ , \label{04.01}%
\end{equation}
where $\phi_{\pi}\left(  p\right)  $ is given by
\begin{equation}
\phi_{\pi}\left(  p\right)  =-i\,2\,g_{0}G_{0}\left(  p\right)  \int
\frac{d^{4}p^{\prime}}{\left(  2\pi\right)  ^{4}}G_{0}\left(  p^{\prime
}\right)  \,\mathbb{T}\text{r}\left(  i\,\gamma_{5}\tau^{i}i\,S\left(
p^{\prime}+\frac{1}{2}P\right)  \,i\,\gamma_{5}\,\tau^{i}\,\phi_{\pi}\left(
p^{\prime}\right)  \,i\,S\left(  p^{\prime}-\frac{1}{2}P\right)  \right)
\label{04.02BSPion0}%
\end{equation}
which is represented in Fig.~\ref{FigBetheSalpeter}. Note that in equation
(\ref{04.02BSPion0}) there is no summation with respect to the isospin index
$i$.

The solution of equation (\ref{04.02BSPion0}) is straightforward and gives
\begin{equation}
\phi_{\pi}\left(  p\right)  =g_{\pi qq}G_{0}\left(  p\right)  \ \ .
\label{04.03}%
\end{equation}
The pion mass is obtained from the BSE, which can be easily rewritten in terms
of the pseudo-scalar polarizability
\begin{equation}
\delta^{ij}\Pi_{PS}\left(  P^{2}\right)  =-i\int\frac{d^{4}p^{\prime}}{\left(
2\pi\right)  ^{4}}G_{0}^{2}\left(  p^{\prime}\right)  \,\mathbb{T}%
\text{r}\left(  i\,\gamma_{5}\tau^{i}i\,S\left(  p^{\prime}+\frac{1}%
{2}P\right)  \,i\,\gamma_{5}\tau^{j}i\,S\left(  p^{\prime}-\frac{1}%
{2}P\right)  \right)  \label{04.04PSpolarization}%
\end{equation}
and equation (\ref{04.02BSPion0}) becomes
\begin{equation}
1=2\,g_{0}\,\Pi_{PS}\left(  P^{2}=m_{\pi}^{2}\right)  \ \ .
\label{04.05MassBoundState}%
\end{equation}
The normalization constant $g_{\pi qq}$ is obtained by the usual normalization
condition of the BSE which can be rewritten as
\begin{equation}
\frac{1}{g_{\pi qq}^{2}}=-\left(  \frac{\partial\Pi_{PS}}{\partial P^{2}%
}\right)  _{P^{2}=m_{\pi}^{2}}~. \label{04.06NormaBoundState}%
\end{equation}

As shown in table \ref{Table 1} we obtain the physical pion mass for
reasonable values of the current quark mass $m_{0}$.

The model realizes the Goldstone theorem. To see it explicitly we can go to
the exact chiral limit, by choosing the reference frame where $P^{\mu}=\left(
M,\vec{0}\right)  $ and taking the limit $M\rightarrow0.$ The explicit
realization of the Goldstone theorem arises because chiral symmetry implies
that we must use the same kernel in the Dyson equation for the mass, equation
(\ref{03.08}), and in the BSE for the pion, equation (\ref{04.02BSPion0})
\cite{DelbourgoScadron79}. Notice that $G_{p}\left(  p\right)  $\ is not
constrained in the procedure.

\section{The quark-photon vertex.}

\label{SecQuarkPhotonVertex}\setcounter{equation}{0}

In order to study the electromagnetic properties of the pion we describe the
coupling of the dressed quarks to the photon. The usual approach to the quark
photon vertex is to exploit the Ward-Takahashi Identity (WTI), which is a
consequence of gauge invariance in QED. The WTI is satisfied order by order in
perturbation theory and must be satisfied also non perturbatively in order to
have the right normalization for the quark-photon vertex.

The WTI for the fermion-photon vertex is
\begin{equation}
\left(  p_{1}-p_{2}\right)  _{\mu}\Gamma^{\mu}\left(  p_{1},p_{2}\right)
=S^{-1}\left(  p_{1}\right)  -S^{-1}\left(  p_{2}\right)  \ . \label{05.01}%
\end{equation}
This equation constrains only the longitudinal component of the proper vertex
and therefore provides no information on the transverse part of $\Gamma^{\mu
}\left(  p_{1},p_{2}\right)  $.

Ball and Chiu \cite{BallChiu80} have given the form of the most general
fermion-photon vertex that satisfies the WTI. It consists of a
longitudinally-constrained part given by
\begin{align}
\Gamma_{BC}^{\mu}\left(  p_{1},p_{2}\right)   &  =\frac{1}{2}\left[  \frac
{1}{Z\left(  p_{1}\right)  }+\frac{1}{Z\left(  p_{2}\right)  }\right]
\gamma^{\mu}+\frac{1}{2}\left[  \frac{1}{Z\left(  p_{1}\right)  }-\frac
{1}{Z\left(  p_{2}\right)  }\right]  \frac{\left(  p_{1}+p_{2}\right)  ^{\mu}%
}{p_{1}^{2}-p_{2}^{2}}\left(  \not p  _{1}+\not p  _{2}\right) \nonumber\\
&  -\left[  \frac{m\left(  p_{1}\right)  }{Z\left(  p_{1}\right)  }%
-\frac{m\left(  p_{2}\right)  }{Z\left(  p_{2}\right)  }\right]  \frac{\left(
p_{1}+p_{2}\right)  ^{\mu}}{p_{1}^{2}-p_{2}^{2}}\ \ , \label{05.03BallChiu}%
\end{align}
and a transverse part which is described in term of a basis of eight
transverse vectors $T_{i}^{\mu}\left(  p_{1},p_{2}\right)  $ given in appendix
\ref{AppQuarkPhotonVertex}. The full quark-photon vertex can be written as
\begin{equation}
\Gamma^{\mu}\left(  p_{1},p_{2}\right)  =\Gamma_{BC}^{\mu}\left(  p_{1}%
,p_{2}\right)  +\sum_{i=1}^{8}\mathcal{V}_{i}\left(  p_{1},p_{2}\right)
T_{i}^{\mu}\left(  p_{1},p_{2}\right)  \ , \label{05.04quarkphotonvertex}%
\end{equation}
where $\mathcal{V}_{i}\left(  p_{1},p_{2}\right)  $ are not constrained scalar
functions of $p_{1}$ and $p_{2}$ with the correct C, P, T invariance properties.

In our scheme the coupling of photons to quarks arises automatically by
implementing gauge invariance in our lagrangian. The usual way to proceed is
to introduce path ordered exponentials in the definition of the currents and
therefore they become
\begin{align}
J_{S}\left(  x\right)   &  =\int d^{4}y\,G_{0}\left(  y\right)  \bar{\psi
}\left(  x+\frac{1}{2}y\right)  ~\mathcal{P}\left(  e^{-iQ\int_{x-\frac{1}%
{2}y}^{x+\frac{1}{2}y}dz^{\mu}A_{\mu}\left(  z\right)  }\right)  ~\psi\left(
x-\frac{1}{2}y\right)  \ ,\label{05.07aScalarGaugeICurrent}\\[0.75cm]
\vec{J}_{5}\left(  x\right)   &  =\int d^{4}y\,G_{0}\left(  y\right)
\bar{\psi}\left(  x+\frac{1}{2}y\right)  ~\mathcal{P}\left(  e^{-iQ\int
_{x}^{x+\frac{1}{2}y}dz^{\mu}A_{\mu}\left(  z\right)  }\right)  ~i\vec{\tau
}\gamma_{5}~\mathcal{P}\left(  e^{-iQ\int_{x-\frac{1}{2}y}^{x}dz^{\mu}A_{\mu
}\left(  z\right)  }\right)  ~\psi\left(  x-\frac{1}{2}y\right)
\ ,\label{05.07bPseudoscalarGaugeICurrent}\\[0.75cm]
J_{p}\left(  x\right)   &  =\int d^{4}y\,G_{p}\left(  y\right)  \frac{1}%
{2}\left[  \bar{\psi}\left(  x+\frac{1}{2}y\right)  ~\mathcal{P}\left(
e^{-iQ\int_{x-\frac{1}{2}y}^{x+\frac{1}{2}y}dz^{\mu}A_{\mu}\left(  z\right)
}\right)  ~i\ \rlap{$/$}\hspace{-0.05cm}D\ \psi\left(  x-\frac{1}{2}y\right)
\right. \nonumber\\
&  ~~~~~~~~~~~~~~~~~~~~-\left.  i\bar{\psi}\left(  x+\frac{1}{2}y\right)
\overleftarrow{\rlap{$/$}\hspace*{-0.05cm}D}~\mathcal{P}\left(  e^{-iQ\int
_{x-\frac{1}{2}y}^{x+\frac{1}{2}y}dz^{\mu}A_{\mu}\left(  z\right)  }\right)
~\psi\left(  x-\frac{1}{2}y\right)  \right]  \ ,
\label{05.07cMomentumGaugeICurrent}%
\end{align}
where the quark charge is $Q=e\left(  \vec{\tau}.\hat{n}+1/3\right)  /2$ with
$\hat{n}=\left(  0,0,1\right)  ,$ $\rlap{$/$}\hspace*{-0.05cm}D\psi\left(
x\right)  =\rlap{$/$}\hspace*{-0.03cm}\partial\psi\left(  x\right)
+iQ\rlap{$/$}\hspace*{-0.05cm}A\left(  x\right)  \psi\left(  x\right)  $ and
$\bar{\psi}\left(  x\right)  \overleftarrow{\rlap{$/$}\hspace*{-0.05cm}%
D}=\partial^{\mu}\bar{\psi}\left(  x\right)  \gamma_{\mu}-i\bar{\psi}\left(
x\right)  \rlap{$/$}\hspace*{-0.05cm}A\left(  x\right)  Q.$ In this way
$J_{S}$ and $J_{p}$ and $\vec{J}_{5}^{\,2}$\ become invariant under local
gauge transformations.

The evaluation of the $z_{\mu}$ integrals in Eqs.
(\ref{05.07aScalarGaugeICurrent}-\ref{05.07cMomentumGaugeICurrent}) implies a
choice of path. The path dependence is implicit in the non locality of the
interaction and cannot be avoided. The difference of the contribution between
two paths is a gauge invariant quantity which is associated with the magnetic
flux through any closed surface defined by the two paths. However, when the
photon momentum vanishes the path dependence disappears.

\begin{figure}[ptb]
\begin{center}
\includegraphics[height=3.9104cm,width=3.6446cm]
{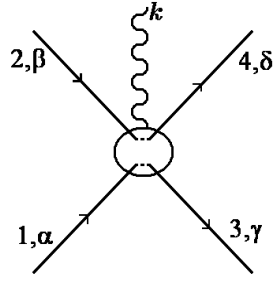}
\end{center}
\caption{Four quark- photon vertex $\Gamma_{i,\mu}^{(4q\gamma)}\left(
p_{1},p_{2};p_{3},p_{4}\right)  $ where $i=S,5,p$ recalls the current from
which the vertex originates}%
\label{Fig4quarkphoton}%
\end{figure}

\begin{figure}[ptb]
\begin{center}
\includegraphics[height=2.8298cm,width=8.9165cm]
{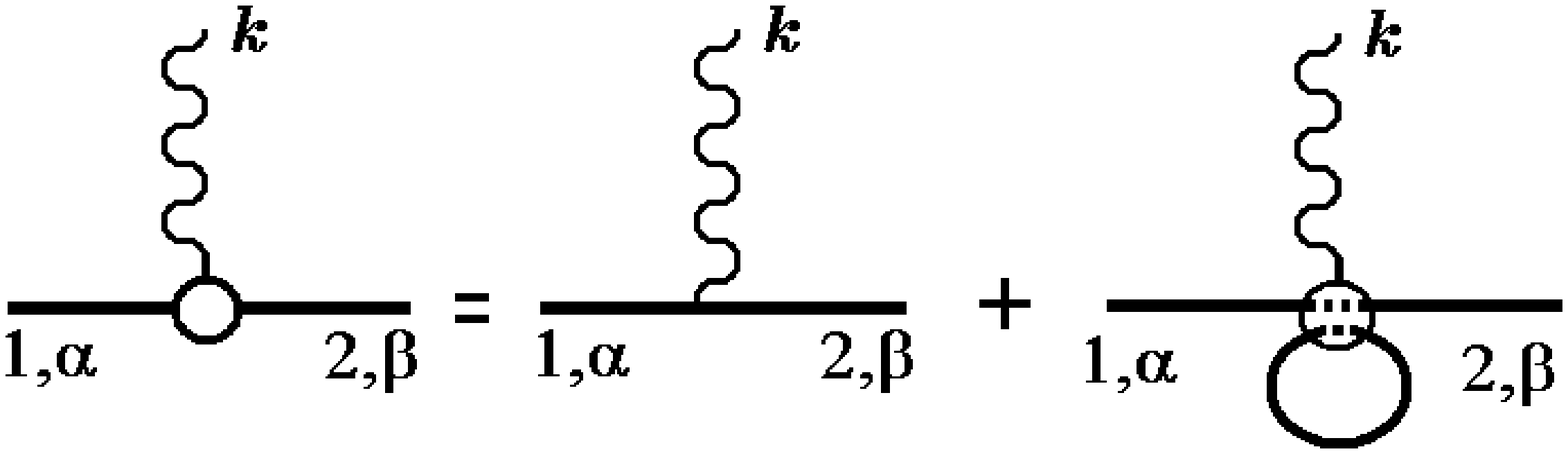}
\end{center}
\caption{Diagramatic representation of eq. (\ref{05.09DressedQuarkPhoton0}).}%
\label{FigQuarkPhoton0}%
\end{figure}

The quantization of the photon field in Eqs.(\ref{05.07aScalarGaugeICurrent}
-\ref{05.07cMomentumGaugeICurrent}) leads to the new vertex shown in
Fig.~\ref{Fig4quarkphoton}, whose contribution has been fully worked out in
Appendix \ref{AppQuarkPhotonVertex}. The full quark-photon vertex can be
constructed in two steps. The first consists in the renormalization of the
bare quark-photon vertex by the 4 quarks one photon vertex. Let us call the
new vertex, shown in Fig.~\ref{FigQuarkPhoton0}, $\Gamma_{0}^{\mu}\left(
p_{1},p_{2}\right)  $. Using Eqs.(\ref{0A.12Scalar4q-fVertex}) and
(\ref{0A.22Momento4q-fVertex}) of Appendix \ref{AppQuarkPhotonVertex}\ we get
\begin{align}
\Gamma_{0}^{\mu}\left(  p_{1},p_{2}\right)   &  =\gamma^{\mu}-\alpha
_{0}\left[  \left(  p_{1}+p_{2}\right)  ^{\mu}\,\mathbb{V}_{0a}\left(  \bar
{p},k\right)  +k^{\mu}\,\mathbb{V}_{0b}\left(  \bar{p},k\right)  \right]
-\nonumber\\
&  \alpha_{p}\frac{1}{2}\left[  G_{p}\left(  p_{1}\right)  +G_{p}\left(
p_{2}\right)  \right]  \mathbb{\gamma}^{\mu}-\nonumber\\
&  \alpha_{p}\frac{\rlap{$/$}p_{1}+\rlap{$/$}p_{2}}{2}\left[  \left(
p_{1}+p_{2}\right)  ^{\mu}\,\mathbb{V}_{pa}\left(  \bar{p},k\right)  +k^{\mu
}\,\mathbb{V}_{pb}\left(  \bar{p},k\right)  \right]
\label{05.09DressedQuarkPhoton0}%
\end{align}
with $k=p_{2}-p_{1},$ $\bar{p}=\frac{p_{1}+p_{2}}{2}$ and $\mathbb{V}_{0a},$
$\mathbb{V}_{0b},$ $\mathbb{V}_{pa}$ and $\mathbb{V}_{pb}$ given in Eqs
(\ref{0A.11}) and (\ref{0A.21}) of Appendix \ref{AppQuarkPhotonVertex}. Both
the scalar and momentum currents contribute to this vertex.

\begin{figure}[ptb]
\begin{center}
\includegraphics[height=2.1066cm,width=9.6659cm]
{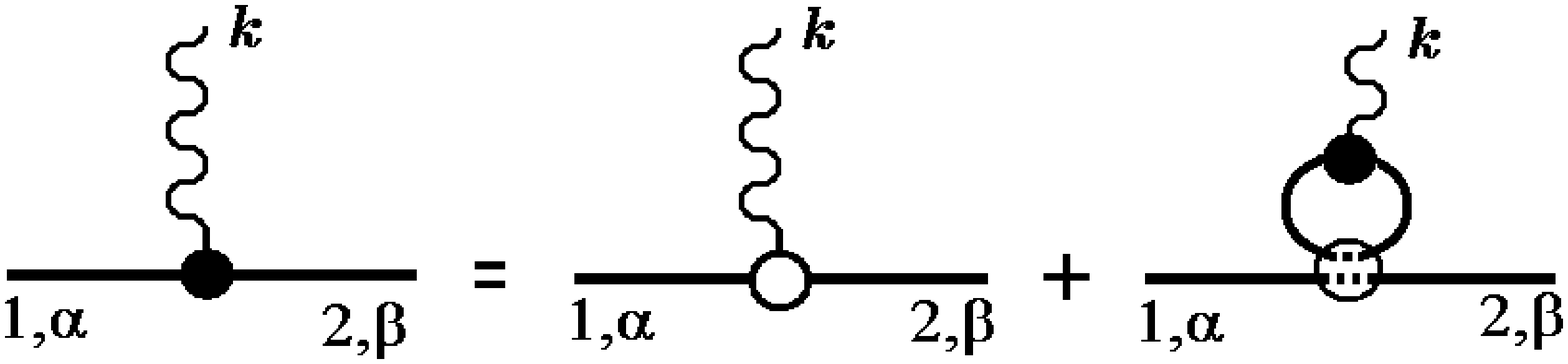}
\end{center}
\caption{Diagramatic representation of eq. (\ref{05.10DSE-QuarkPhoton}).}%
\label{FigQuarkPhoton1}%
\end{figure}

The second step consists in the insertion of $\Gamma_{0}^{\mu}\left(
p_{1},p_{2}\right)  $\ in the equation for the quark-photon vertex,
represented in Fig.~\ref{FigQuarkPhoton1}, which produces the dressed
quark-photon vertex,
\begin{gather}
i\Gamma^{\mu}\left(  p_{1},p_{2}\right)  =i\Gamma_{0}^{\mu}\left(  p_{1}%
,p_{2}\right)  +i2g_{0}G_{0}\left(  \frac{p_{1}+p_{2}}{2}\right)  \int
\frac{d^{4}p}{\left(  2\pi\right)  ^{4}}G_{0}\left(  p\right)  \,\mathbb{T}%
\text{r}\left[  i\,S\left(  p-\frac{k}{2}\right)  \,i\,S\left(  p+\frac{k}%
{2}\right)  i\Gamma^{\mu}\left(  p-\frac{k}{2},p+\frac{k}{2}\right)  \right]
\nonumber\\
+i2g_{p}~\left(  \frac{\rlap{$/$}p_{1}+\rlap{$/$}p_{2}}{2}\right)
~G_{p}\left(  \frac{p_{1}+p_{2}}{2}\right)  \int\frac{d^{4}p}{\left(
2\pi\right)  ^{4}}G_{p}\left(  p\right)  \,\mathbb{T}\text{r}\left[
i\,S\left(  p-\frac{k}{2}\right)  \rlap{$/$}p\,i\,S\left(  p+\frac{k}%
{2}\right)  i\Gamma^{\mu}\left(  p-\frac{k}{2},p+\frac{k}{2}\right)  \right]
~~. \label{05.10DSE-QuarkPhoton}%
\end{gather}
It is easy to show that the solution of equation (\ref{05.10DSE-QuarkPhoton})
is just
\begin{equation}
\Gamma^{\mu}\left(  p_{1},p_{2}\right)  =\Gamma_{0}^{\mu}\left(  p_{1}%
,p_{2}\right)  ~~. \label{05.11}%
\end{equation}

In agreement with our previous discussion, the quark-photon vertex can be
rewritten in the form given by equation (\ref{05.04quarkphotonvertex}),
\begin{equation}
\Gamma^{\mu}\left(  p_{1},p_{2}\right)  =\Gamma_{BC}^{\mu}\left(  p_{1}%
,p_{2}\right)  +\mathcal{V}_{1}\left(  p_{1},p_{2}\right)  T_{1}^{\mu}\left(
p_{1},p_{2}\right)  +\mathcal{V}_{2}\left(  p_{1},p_{2}\right)  T_{2}^{\mu
}\left(  p_{1},p_{2}\right)  \ , \label{05.12DressedQuarkPhoton}%
\end{equation}
where
\begin{align}
\mathcal{V}_{1}\left(  p_{1},p_{2}\right)   &  =\frac{2\alpha_{0}}{p_{2}%
^{2}-p_{1}^{2}}\mathbb{V}_{0b}\left(  \bar{p},k\right)  \ ,\label{05.13aa}\\
\mathcal{V}_{2}\left(  p_{1},p_{2}\right)   &  =\frac{2\alpha_{p}}{p_{2}%
^{2}-p_{1}^{2}}\mathbb{V}_{pb}\left(  \bar{p},k\right)  \ . \label{05.13bb}%
\end{align}
In summary, we have constructed a dressed quark-photon vertex which has all
the desired properties. It is gauge invariant in the local sense, without
being obtained from the WTI. Two new terms appear from the restoration of the
local gauge symmetry. The simplest expressions for these terms can be built
assuming that a straight line joins the two points characterizing the non
local currents. For that simple input we obtain
\begin{align}
\mathcal{V}_{1}\left(  p_{1},p_{2}\right)   &  =\frac{1}{p_{2}^{2}-p_{1}^{2}%
}\int_{-1}^{1}d\lambda\;\lambda\;\left[  \frac{d}{dp^{2}}\frac{m\left(
p\right)  }{Z\left(  p\right)  }\right]  _{p^{2}=\left(  \bar{p}-\frac
{\lambda}{2}k\right)  ^{2}}\ ,\label{05.161a}\\
\mathcal{V}_{2}\left(  p_{1},p_{2}\right)   &  =\frac{-1}{p_{2}^{2}-p_{1}^{2}%
}\int_{-1}^{1}d\lambda\;\lambda\;\left[  \frac{d}{dp^{2}}\frac{1}{Z\left(
p\right)  }\right]  _{p^{2}=\left(  \bar{p}-\frac{\lambda}{2}k\right)  ^{2}%
}\ , \label{05.161b}%
\end{align}
which can be directly evaluated from the form of the quark propagator.

\section{The axial current and $f_{\pi}$.}

\label{SecAxialCurrent_fpi}\setcounter{equation}{0}Lets us now turn to the
coupling of the axial current to quarks. The WTI in this case is
\begin{equation}
\left(  p_{2}-p_{1}\right)  _{\mu}\Gamma_{5}^{\mu}\left(  p_{1},p_{2}\right)
\tau^{i}=S^{-1}\left(  p_{2}\right)  \gamma_{5}\tau^{i}+\gamma_{5}\tau
^{i}S^{-1}\left(  p_{1}\right)  -2m_{0}i~\Gamma_{5}\left(  p_{1},p_{2}\right)
\tau^{i}\ . \label{05a.01}%
\end{equation}
The last term, proportional to $m_{0},$ disappears in the chiral limit.

Our first step is to obtain $\Gamma_{5}\left(  p_{1},p_{2}\right)  \tau^{i},$
which corresponds to the dressing of the vertex $i\gamma_{5}\tau^{i}.$ Now the
external probe is a pseudoscalar-isovector therefore we will have contribution
from the pseudoscalar current, $\vec{J}_{5},$%
\begin{gather}
\Gamma_{5}\left(  p_{1},p_{2}\right)  \tau^{i}=i\gamma_{5}\tau^{i}%
+i2g_{0}G_{0}\left(  \bar{p}\right)  \,i\gamma_{5}\tau^{j}\,\nonumber\\
\int\frac{d^{4}p}{\left(  2\pi\right)  ^{4}}G_{0}\left(  p\right)  \,\left(
-\right)  \mathbb{T}\text{r}\left[  i\,S\left(  p-\frac{k}{2}\right)
\,i\gamma_{5}\tau^{j}\,i\,S\left(  p+\frac{k}{2}\right)  \Gamma_{5}\left(
p-\frac{k}{2},p+\frac{k}{2}\right)  \tau^{i}\right]  ~~. \label{05a.02}%
\end{gather}
with $p_{1}=\bar{p}-\frac{k}{2}$ and $p_{2}=\bar{p}+\frac{k}{2}.$\ This
equation can be easily solved obtaining%
\begin{equation}
\Gamma_{5}\left(  p_{1},p_{2}\right)  =i\gamma_{5}\left[  1+2g_{0}%
~G_{0}\left(  \bar{p}\right)  \frac{F_{0}\left(  k^{2}\right)  }{1-2g_{0}%
~\Pi_{PS}\left(  k^{2}\right)  }\right]  \label{05a.04}%
\end{equation}
where $F_{0}\left(  k^{2}\right)  $ is defined by%
\begin{equation}
\delta^{ij}F_{0}\left(  k^{2}\right)  =-i\int\frac{d^{4}p}{\left(
2\pi\right)  ^{4}}G_{0}\left(  p\right)  \,\mathbb{T}\text{r}\left(
i\,\gamma_{5}\tau^{i}i\,S\left(  p-\frac{1}{2}k\right)  \,i\,\gamma_{5}%
\tau^{j}i\,S\left(  p+\frac{1}{2}k\right)  \right)  ~~. \label{05a.03}%
\end{equation}

We will obtain the dressed vertex $\Gamma_{5}^{\mu}\left(  p_{1},p_{2}\right)
$ applying to the $W^{\pm}$ bosons the same procedure developed for photons in
the previous section. The explicit calculation is given in Appendix
\ref{AppQuarkWVertex}. The final result, given in equation (\ref{0B.25}), can
be rewritten in the following way:%
\begin{align}
\Gamma_{5}^{\mu}\left(  p_{1},p_{2}\right)   &  =\tilde{\Gamma}_{5}^{\mu
}\left(  p_{1},p_{2}\right) \nonumber\\
&  +\mathcal{A}_{1}\left(  p_{1},p_{2}\right)  T_{1}^{\mu}\left(  p_{1}%
,p_{2}\right)  \gamma_{5}+\mathcal{V}_{2}\left(  p_{1},p_{2}\right)
T_{2}^{\mu}\left(  p_{1},p_{2}\right)  \gamma_{5}~~, \label{05a.10}%
\end{align}
where%
\begin{equation}
\mathcal{A}_{1}\left(  p_{1},p_{2}\right)  =\frac{2\alpha_{0}}{p_{2}^{2}%
-p_{1}^{2}}\left[  \frac{2G_{0}\left(  \bar{p}\right)  -G_{0}\left(
p_{1}\right)  -G_{0}\left(  p_{2}\right)  }{\left(  p_{1}-p_{2}\right)  ^{2}%
}+\mathbb{A}_{0b}\left(  \bar{p},k\right)  \right]  \ , \label{05a.11}%
\end{equation}
with $\mathbb{A}_{0b}\left(  \bar{p},k\right)  $\ given in eq (\ref{0A.16d}),
and%
\begin{align}
\tilde{\Gamma}_{5}^{\mu}\left(  p_{1},p_{2}\right)   &  =\frac{1}{2}\left[
\frac{1}{Z\left(  p_{1}\right)  }+\frac{1}{Z\left(  p_{2}\right)  }\right]
\gamma^{\mu}\gamma_{5}+\frac{1}{2}\left[  \frac{1}{Z\left(  p_{2}\right)
}-\frac{1}{Z\left(  p_{1}\right)  }\right]  \frac{\left(  p_{1}+p_{2}\right)
^{\mu}}{p_{2}^{2}-p_{1}^{2}}\left(  \not p  _{1}+\not p  _{2}\right)
\gamma_{5}\nonumber\\
&  -\left[  \frac{m\left(  p_{1}\right)  }{Z\left(  p_{1}\right)  }%
+\frac{m\left(  p_{2}\right)  }{Z\left(  p_{2}\right)  }\right]  \frac{\left(
p_{2}-p_{1}\right)  ^{\mu}}{\left(  p_{2}-p_{1}\right)  ^{2}}\gamma
_{5}-i2m_{0}\Gamma_{5}\left(  p_{1},p_{2}\right)  \frac{\left(  p_{2}%
-p_{1}\right)  ^{\mu}}{\left(  p_{2}-p_{1}\right)  ^{2}}, \label{05a.12}%
\end{align}
is the longitudinally constrained part of $\Gamma_{5}^{\mu}\left(  p_{1}%
,p_{2}\right)  .$%

\begin{figure}
[ptb]
\begin{center}
\includegraphics[
height=3.8551cm,
width=7.2379cm
]%
{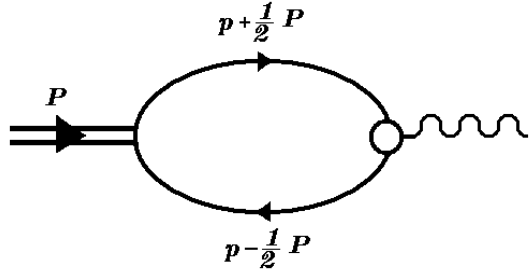}%
\caption{Diagramatic representation of the pion decay process corresponding to
eq. (\ref{05a.13}).}%
\label{FigFpi0}%
\end{center}
\end{figure}

Let us now to look for the pion decay constant. In this calculation only the
longitudinal current is needed and so we do not need to choose a particular
path. There are several equivalent ways for obtaining the pion decay constant,
depending where the quark-quark interaction is included. The simplest one is
through the diagram depicted in figure \ref{FigFpi0}, in which all the
$q\bar{q}$ bubbles are included in the Bethe-Salpeter pion amplitude. The
axial vertex includes those vertex corrections which can not be included in
the bound state amplitude, which in our case corresponds to those depicted in
Fig \ref{FigQuarkPhoton0}. Therefore, the pion decay constant is defined by,%
\begin{gather}
i~\delta_{i,j}~f_{\pi}~P^{\mu}=\nonumber\\
\int\frac{d^{4}p}{\left(  2\pi\right)  ^{4}}\left(  -\right)  \mathbb{T}%
\text{r}\left[  i\,\phi_{\pi}\left(  p,P\right)  \ i\gamma_{5}\ \tau
^{i}\ i\ S\left(  p-\frac{1}{2}P\right)  \ \Gamma_{5,0}^{\mu}\left(
p+\frac{1}{2}P,p-\frac{1}{2}P\right)  \ \frac{\tau^{j}}{2}\ i\ S\left(
p+\frac{1}{2}P\right)  \right]  ~, \label{05a.13}%
\end{gather}
where $\Gamma_{5,0}^{\mu}\left(  p_{1},p_{2}\right)  $ is given in equation
(\ref{0B.20}) of Appendix \ref{AppQuarkWVertex} and $\phi_{\pi}\left(
p,P\right)  $ given by equation (\ref{04.03}). A direct calculation gives the
result,%
\begin{equation}
f_{\pi}=\frac{g_{\pi qq}}{P^{2}}\left\{  \frac{1}{2}\left(  1-2g_{0}\Pi
_{PS}\left(  P^{2}\right)  \right)  F_{1}\left(  P^{2}\right)  +m_{0}%
F_{0}\left(  P^{2}\right)  \right\}  ~~. \label{05a.22}%
\end{equation}
with $F_{0}$ given in equation (\ref{05a.03}) and
\begin{equation}
F_{1}\left(  P^{2}\right)  =-\int\frac{d^{4}p}{\left(  2\pi\right)  ^{4}%
}\left(  G_{0}\left(  p+\frac{P}{2}\right)  +G_{0}\left(  p-\frac{P}%
{2}\right)  \right)  \mathbb{T}\text{r}\left[  i\,S\left(  p\right)  \right]
~. \label{05a.23}%
\end{equation}

We have two clearly distinguishable situations. In the chiral limit $f_{\pi}$
is determined by the term with $F_{1}\left(  P^{2}\right)  $. In this limit,
$F_{1}\left(  0\right)  $ is the integral evaluated in equation (\ref{03.08}).
Using in the latter equations (\ref{04.05MassBoundState}) and
(\ref{04.06NormaBoundState}) and substituting, we obtain%
\begin{equation}
f_{\pi}=\frac{\alpha_{0}}{g_{\pi qq}}=\frac{m\left(  0\right)  }{g_{\pi
qq}Z\left(  0\right)  }~. \label{05a.24}%
\end{equation}
The last form of $f_{\pi}$ is the Goldberger-Treiman relation. The crucial
point for obtaining this result is the chiral symmetric structure of the
interaction, which connects the kernel present in the calculation of $f_{\pi}%
$, associated with the pseudoscalar-isovector current present in our
lagrangian, with the one present in the evaluation of the mass term $m\left(
0\right)  ,$ associated to the scalar-isoscalar current.

If we work with physical pions ($m_{0}\neq0),$ the pion decay constant is
given by%
\begin{equation}
f_{\pi}=\frac{m_{0}}{m_{\pi}^{2}}g_{\pi qq}F_{0}\left(  m_{\pi}^{2}\right)  ~.
\label{05a.25}%
\end{equation}
In this case the surviving contribution arises from the pion pole present in
$\Gamma_{5}\left(  p_{1},p_{2}\right)  $.

Using equations (\ref{05a.03}), (\ref{03.07Mass}) and (\ref{03.11}) we can
obtain approximative expressions for $F_{0}\left(  m_{\pi}^{2}\right)  $ which
lead directly to the Gell-Mann-Oakes-Renner relation%
\begin{equation}
f_{\pi}^{2}=-N_{f}~\left\langle \bar{q}q\right\rangle \frac{m_{0}}{m_{\pi}%
^{2}}+\mathcal{O}\left(  m_{0},m_{\pi}^{2}\right)  ~. \label{05a.27}%
\end{equation}

We have seen that we have different expressions for $f_{\pi}$ in the chiral
limit and in the physical case. They originate from different terms of the
axial current. Nevertheless, chiral symmetry guarantees that (\ref{05a.24}) is
the limit of equation (\ref{05a.25}) when $m_{0}$ goes to zero.

In table \ref{Table 1} we give numerical values for $f_{\pi}$ for scenarios S1
and S2 previously discussed. They are in reasonably good agreement with the
experimental results. Moreover, this two scenarios are describing the same
physics, even if they have a very different origin.

From reference \cite{MarisRobertsTandy98} we can infer that the pion decay
constant is determined using equation (\ref{05a.13}) with the approximated
expression for the axial vertex,%
\begin{equation}
\Gamma_{5}^{\prime\mu}\left(  p_{1},p_{2}\right)  =\frac{1}{2}\left[  \frac
{1}{Z\left(  p_{1}\right)  }+\frac{1}{Z\left(  p_{2}\right)  }\right]
\gamma^{\mu}\gamma_{5}+\frac{1}{2}\left[  \frac{1}{Z\left(  p_{1}\right)
}-\frac{1}{Z\left(  p_{2}\right)  }\right]  \frac{\left(  p_{1}+p_{2}\right)
^{\mu}}{p_{1}^{2}-p_{2}^{2}}\left(  \rlap{$/$}p_{1}+\rlap{$/$}p_{2}\right)
\gamma_{5}+\mathcal{O}\left(  p_{1}-p_{2}\right)  \ .\label{05a.31}%
\end{equation}
In this expression $\mathcal{O}\left(  p_{1}-p_{2}\right)  $\ implies that
$\left(  p_{1}-p_{2}\right)  _{\mu}\Gamma_{5}^{\prime\mu}\left(  p_{1}%
,p_{2}\right)  $ is determined up to terms of order $\left(  p_{1}%
-p_{2}\right)  ^{2}=P^{2}.$ This expression is the minimal generalization of
the bare $\gamma^{\mu}\gamma_{5}$ vertex in the case where $Z\left(  p\right)
\neq1.$ It is straightforward to proof that the $f_{\pi}$ obtained in this way
differs from the exact one by corrections of order $m_{\pi}^{2}$. Therefore,
equation (\ref{05a.31}) gives the right result in the chiral limit and a good
approximation to the exact value in the physical case. The
Gell-Mann-Oakes-Renner relation is also well reproduced in this approximation.
Nevertheless, equation (\ref{05a.31}) is not a good expression for the axial
vertex, for instance the pion pole is not present in this expression. So we
conclude that $\Gamma_{5}^{\prime\mu}\left(  p_{1},p_{2}\right)  $ is a good
approximation of the longitudinal part of $\Gamma_{5,0}^{\mu}\left(
p_{1},p_{2}\right)  $ in the vicinity of the pion mass ($P^{2}\sim m_{\pi}%
^{2})$, as it can be see from equation (\ref{0B.26}).

Let us now consider the coupling of the axial current to a line of quarks
through the vertex, $\Gamma_{5}^{\mu}\left(  p_{1},p_{2}\right)  .$ This
coupling contains the direct coupling and the pion pole contribution. We
observe from equations (\ref{05a.12}) and (\ref{05a.04}) that the pion pole
contribution goes trough the $\Gamma_{5}\left(  p_{1},p_{2}\right)  $ term
when the chiral symmetry is explicitly broken. In the chiral limit, the pion
pole is present in the term with quark masses in $\tilde{\Gamma}_{5}^{\mu
}\left(  p_{1},p_{2}\right)  $. The full $\Gamma_{5}^{\mu}\left(  p_{1}%
,p_{2}\right)  $ has some dependence on the choice of path. Nevertheless, the
overall procedure preserves all the symmetries and, in particular, gauge symmetry.

The longitudinal part of the axial current, $\tilde{\Gamma}_{5}^{\mu}\left(
p_{1},p_{2}\right)  ,$ is path independent. Using equations (\ref{05a.25}) and
(\ref{05a.27}), and assuming that $P^{2}\lesssim m_{\pi}^{2},$ it can be
written\ in terms of the pion wave function as%
\begin{align}
\tilde{\Gamma}_{5}^{\mu}\left(  p_{1},p_{2}\right)  \frac{\tau^{i}}{2} &
=\frac{1}{2}\left[  \frac{1}{Z\left(  p_{1}\right)  }+\frac{1}{Z\left(
p_{2}\right)  }\right]  \gamma^{\mu}\gamma_{5}\frac{\tau^{i}}{2}+\frac{1}%
{2}\left[  \frac{1}{Z\left(  p_{2}\right)  }-\frac{1}{Z\left(  p_{1}\right)
}\right]  \frac{\left(  p_{1}+p_{2}\right)  ^{\mu}}{p_{2}^{2}-p_{1}^{2}%
}\left(  \not p_{1}+\not p_{2}\right)  \gamma_{5}\frac{\tau^{i}}{2}\nonumber\\
&  -\left[  \frac{m\left(  p_{1}\right)  }{Z\left(  p_{1}\right)  }%
+\frac{m\left(  p_{2}\right)  }{Z\left(  p_{2}\right)  }\right]  \frac{\left(
p_{2}-p_{1}\right)  ^{\mu}}{\left(  p_{2}-p_{1}\right)  ^{2}}\gamma_{5}%
\frac{\tau^{i}}{2}+\nonumber\\
&  \left(  m_{0}+m_{\pi}^{2}f_{\pi}\frac{\phi_{\pi}\left(  p\right)  }%
{P^{2}-m_{\pi}^{2}}\right)  \gamma_{5}\tau^{i}\frac{\left(  p_{2}%
-p_{1}\right)  ^{\mu}}{\left(  p_{2}-p_{1}\right)  ^{2}},\label{05a.33}%
\end{align}
with $p_{1,2}=p\pm\frac{1}{2}P.$ Equation (\ref{05a.33}) manifests the pion
pole explicitly, as is predicted by PCAC.

\section{Electromagnetic Pion form factor.}

\label{SecPionFF}\setcounter{equation}{0}

\begin{figure}[ptb]
\begin{center}
\includegraphics[height=1.5757in,width=5.988in]
{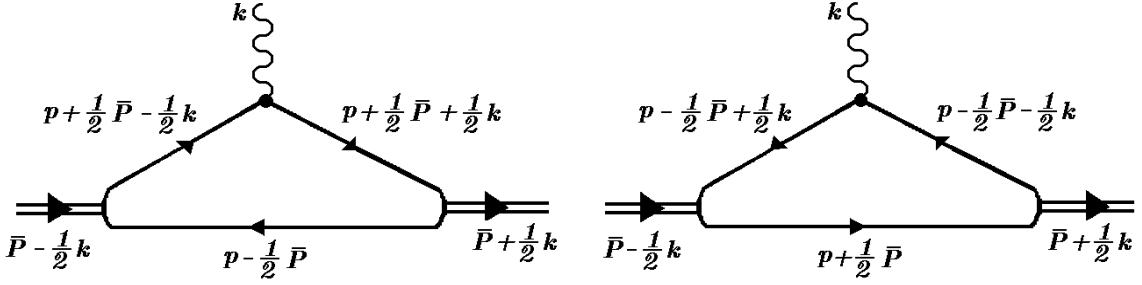}
\end{center}
\caption{Triangle diagram contributing to the pion form factor.}%
\label{FigPionFF1}%
\end{figure}

We begin by considering the triangle diagram of Fig.~\ref{FigPionFF1}. In
order to fix ideas let us consider the interaction of a photon with a $\pi
^{+}$. With the momenta defined as in the figure we have
\begin{gather}
i\,e\,2\,\bar{P}^{\mu}F^{\left(  2q\gamma\right)  }\left(  k^{2}\right)
=\int\frac{d^{4}p}{\left(  2\pi\right)  ^{4}}\left(  -\right)  \,\mathbb{T}%
\text{r}\left[  i\,\gamma_{5}\,\tau^{+}\phi_{\pi}\left(  p-\frac{1}%
{4}k\right)  i\,S\left(  p-\frac{1}{2}\bar{P}\right)  \right.
\nonumber\\[0.2cm]
\left.  \,i\,\gamma_{5}\,\left(  \tau^{+}\right)  ^{\dag}\,\phi_{\pi}\left(
p+\frac{1}{4}k\right)  \,i\,S\left(  p+\frac{1}{2}\bar{P}+\frac{1}{2}k\right)
\,iQ\,\Gamma^{\mu}\left(  p+\frac{1}{2}\bar{P}-\frac{1}{2}k,p+\frac{1}{2}%
\bar{P}+\frac{1}{2}k\right)  \,i\,S\left(  p+\frac{1}{2}\bar{P}-\frac{1}%
{2}k\right)  \right] \nonumber\\[0.2cm]
+\int\frac{d^{4}p}{\left(  2\pi\right)  ^{4}}\left(  -\right)  \,\mathbb{T}%
\text{r}\left[  i\,\gamma_{5}\,\tau^{+}\,\phi_{\pi}\left(  p+\frac{1}%
{4}k\right)  \,i\,S\left(  p-\frac{1}{2}\bar{P}+\frac{1}{2}k\right)
\,i\,Q\,\Gamma^{\mu}\left(  p-\frac{1}{2}\bar{P}-\frac{1}{2}k,p-\frac{1}%
{2}\bar{P}+\frac{1}{2}k\right)  \right. \nonumber\\[0.2cm]
\left.  i\,S\left(  p-\frac{1}{2}\bar{P}-\frac{1}{2}k\right)  \,i\,\gamma
_{5}\,\left(  \tau^{+}\right)  ^{\dag}\,\phi_{\pi}\left(  p-\frac{1}%
{4}k\right)  \,i\,S\left(  p+\frac{1}{2}\bar{P}\right)  \right]  \ \ ,
\label{06.01FFtriangulo}%
\end{gather}
where $\phi_{\pi}$ is the pion Bethe-Salpeter amplitude given in equation
(\ref{04.03}) and $\Gamma^{\mu}\left(  p_{1},p_{2}\right)  $ is the dressed
quark-photon coupling given in equation (\ref{05.12DressedQuarkPhoton}).

\begin{figure}[ptb]
\begin{center}
\includegraphics[height=1.6881in,width=4.7037in]
{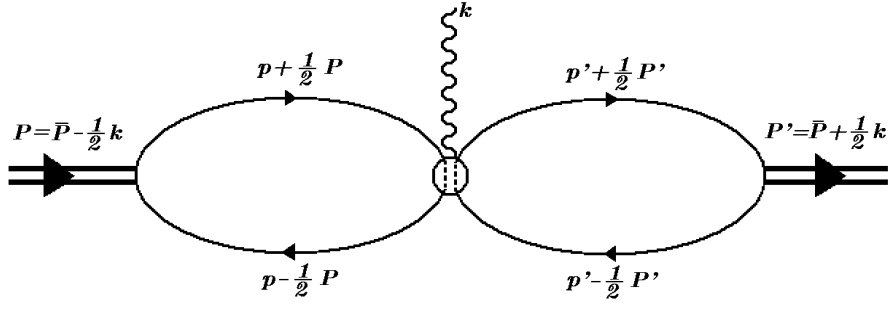}
\end{center}
\caption{Contribution to the form factor coming from the four quark-photon
vertex displayed in Fig. \ref{Fig4quarkphoton}}%
\label{FigPionFF2}%
\end{figure}

There is another contribution to the form factor to be added to the previous
one. The $\vec{J}_{5}^{\,2}\left(  x\right)  $ term produces a four quark
photon vertex evaluated in Appendix \ref{AppQuarkPhotonVertex}\ and given by
equation (\ref{0A.17Pseudo4q-fVertex}). This four quark photon vertex allows
for an additional diagram shown in Fig.~\ref{FigPionFF2}. This contribution
is
\begin{gather}
i\,e\,2\,\bar{P}_{\mu}F^{(4q\gamma)}\left(  k^{2}\right)  =-\int\frac{d^{4}%
p}{\left(  2\pi\right)  ^{4}}\int\frac{d^{4}p^{\prime}}{\left(  2\pi\right)
^{4}}\,\left[  i\,S\left(  p^{\prime}-\frac{1}{2}P^{\prime}\right)
i\,\gamma_{5}\,\left(  \tau^{+}\right)  ^{\dag}\,\phi_{\pi}\left(  p^{\prime
}\right)  i\,S\left(  p^{\prime}+\frac{1}{2}P^{\prime}\right)  \right]
_{\alpha\gamma}\nonumber\\[0.2cm]
\,\left[  \,\Gamma_{5,\mu}^{(4q\gamma)}\left(  p^{\prime}-\frac{1}{2}%
P^{\prime},p^{\prime}+\frac{1}{2}P^{\prime};p+\frac{1}{2}P,p-\frac{1}%
{2}P\right)  \right]  _{\alpha\beta,\gamma\delta}\,\left[  i\,S\left(
p+\frac{1}{2}P\right)  i\,\gamma_{5}\,\tau^{+}\phi_{\pi}\left(  p\right)
i\,S\left(  p-\frac{1}{2}P\right)  \right]  _{\beta\delta}. \label{06.02FF4qf}%
\end{gather}
Therefore, the full electromagnetic form factor is $F\left(  k^{2}\right)
=F^{\left(  2q\gamma\right)  }\left(  k^{2}\right)  +F^{\left(  4q\gamma
\right)  }\left(  k^{2}\right)  $.

Global gauge invariance guarantees the right normalization for the form
factor, $F\left(  k^{2}=0\right)  =1$. For a general case, the right
normalization of the form factor is assured by the Ward identity, equation
(\ref{05.01}), provided that $\phi_{\pi}$ is normalized by equation
(\ref{04.06NormaBoundState}) and we take into account the two contributions
arising from Eqs. (\ref{06.01FFtriangulo}) and (\ref{06.02FF4qf}). When the
kernel used in the pion BSE, equation (\ref{04.05MassBoundState}), is
independent on the total pion momentum, as it is for our models, the
contribution arising from the triangle diagram, equation
(\ref{06.01FFtriangulo}) will assure the correct normalization of the form
factor. Therefore, in our case this property is guaranteed for the full
$\Gamma^{\mu}\left(  p_{1},p_{2}\right)  $ vertex given by equation
(\ref{05.12DressedQuarkPhoton}) and for the Ball-Chiu vertex, $\Gamma
_{BC}^{\mu}\left(  p_{1},p_{2}\right)  ,$ given by equation
(\ref{05.03BallChiu}). In our expression for $\Gamma^{\mu}\left(  p_{1}%
,p_{2}\right)  $ there are additional terms besides $\Gamma_{BC}^{\mu}\left(
p_{1},p_{2}\right)  $ which arise from local gauge invariance. They give
contributions to $F\left(  k^{2}\right)  $ for $k^{2}\neq0$ without modifying
the value at $k^{2}=0.$

Due to the separable nature in $p$ and $p^{\prime}$ and $P$ -independence\ of
the interacting terms in our Lagrangian, one of the integrals in equation
(\ref{06.02FF4qf}) can be done in a trivial manner generating a pion-2
quark-photon vertex. However, when performing the remaining integrals we
realize that this diagram does not contribute to the form factor. This is not
a general result, just a consequence of our particular models. In fact this
term will contribute in a significant way to the parton distribution
\cite{NogueraVento05}. Its contribution is crucial to guarantee isospin
symmetry in the parton distributions and to restore the momentum sum rule.

\begin{figure}[ptb]
\begin{center}
\includegraphics[height=9.3137cm,width=12.3714cm]
{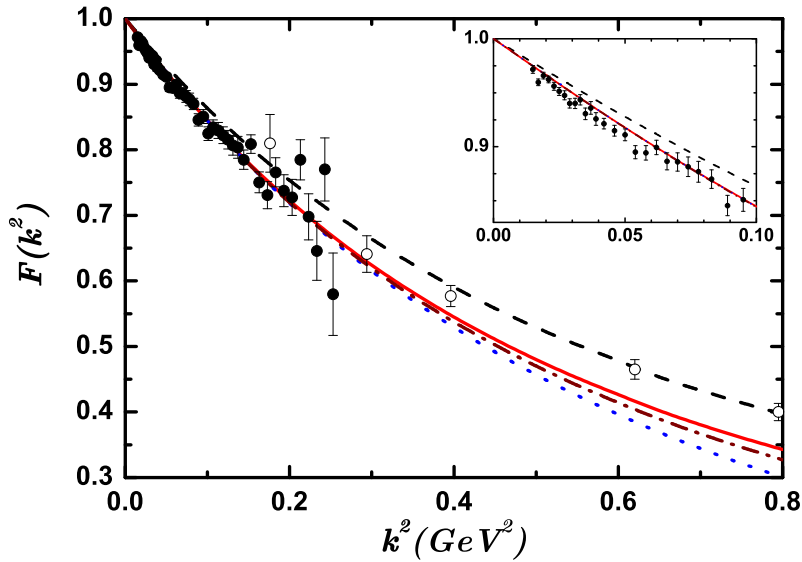}
\end{center}
\caption{Comparison of the Pion form factor calculated in the two defined
scenarios with the Ball-Chiu ansatz and the full vertex of the model.
Dot-dashed curve corresponds to the scenario S1 with Ball-Chiu ansatz; full
curve corresponds to the same scenario with the full vertex. Dotted curve
corresponds to S2 with the Ball-Chiu ansatz; dashed curve corresponds to S2
with the full vertex. Experimental data have been taken from \cite{Amendola86}
(points) and from \cite{Bebek76} (circles).}%
\label{FigFormFactor}%
\end{figure}

We now proceed to a numerical comparison between the calculations, using the
Ball-Chiu vertex, and our full locally gauge invariant vertex, equation
(\ref{05.12DressedQuarkPhoton}), (\ref{05.161a}) and (\ref{05.161b}). We show
in Fig.~\ref{FigFormFactor} the form factor for the two vertices in scenarios
S1 and S2, together with the experimental results \cite{Bebek76,Amendola86}.
We find no important differences for S1 when comparing the Ball-Chiu
prescription to the full vertex. For S2 the correction, which is small for
small $k^{2},$ becomes important for $k^{2}\sim-0.8\operatorname{GeV}^{2}$.
The difference in the calculations arises because in S2 $Z\left(  p\right)
\neq1$. The correction due to $\mathcal{V}_{1}\left(  p_{1},p_{2}\right)  $ is
about 5-7\% at this momentum transfer for both scenarios. The correction due
to $\mathcal{V}_{2}\left(  p_{1},p_{2}\right)  $, present only in S2, is about
24\% for $k^{2}\sim-0.8\operatorname{GeV}^{2}$. Since the two corrections go
in the same direction the overall result changes by about 30\%. We have
confirmed this conclusion introducing a $Z\left(  p\right)  $ different from 1
in case S1.

Regarding the experimental results we observe that the scenario S1 reproduces
well the value of $F\left(  k^{2}\right)  $ for small values of $k$ but
underestimates the form factor for $k^{2}\sim0.8 \operatorname{GeV}^{2}$ by as
much as 12\%. For scenario S2 we observe that the introduction of the full
vertex produces a better description of the form factor for $k^{2}\sim0.8
\operatorname{GeV}^{2}$ but a worst in the small $k$ region. In this last
scenario the difference between the calculated form factor and the
experimental data is always less than 5\%.

The behavior for small $k$ can be analyzed in terms of the pion radius. In
table \ref{Table 1} we give the mean squared radius for the full
electromagnetic vertex and, between brackets, that of the Ball-Chiu
prescription. We observe that the radius is smaller than the experimental
result in all cases. We also observe that the full vertex and the Ball-Chiu
prescription produce the same values since there is no wave function
renormalization in the quark propagator, as in the case S1. In model S2, with
non vanishing wave function renormalization, we observe a difference, of about
15\%, due to the $\mathcal{V}_{2}\left(  p_{1},p_{2}\right)  $ term in
equation (\ref{05.12DressedQuarkPhoton}), confirming our conclusion from the
analysis of the form factors. Summarizing, the use of the full vertex
increases the differences between the calculation and the observation. But
this is not unexpected since no vector mesons have been included in the models
and previous work indicates that this contribution can be of the order of
10-20\% \cite{MarisTandi00, Dorokhov3, OconnellPearceThomasWilliams97}.

\begin{figure}[ptb]
\begin{center}
\includegraphics[height=6.4cm,width=7.8cm]{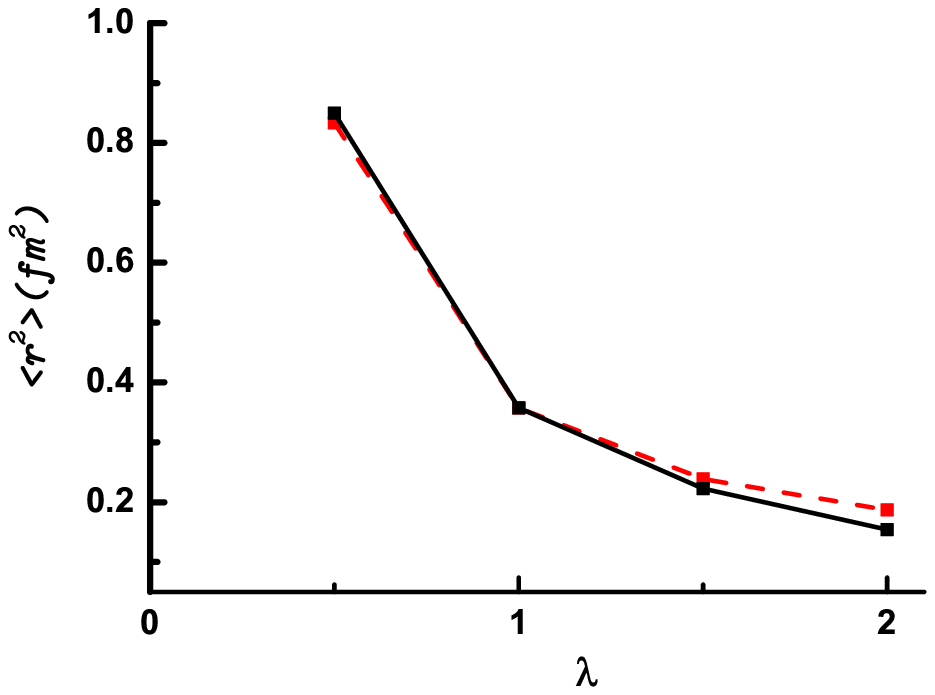}\hspace{2em}
\includegraphics[height=6.4cm,width=7.8cm]{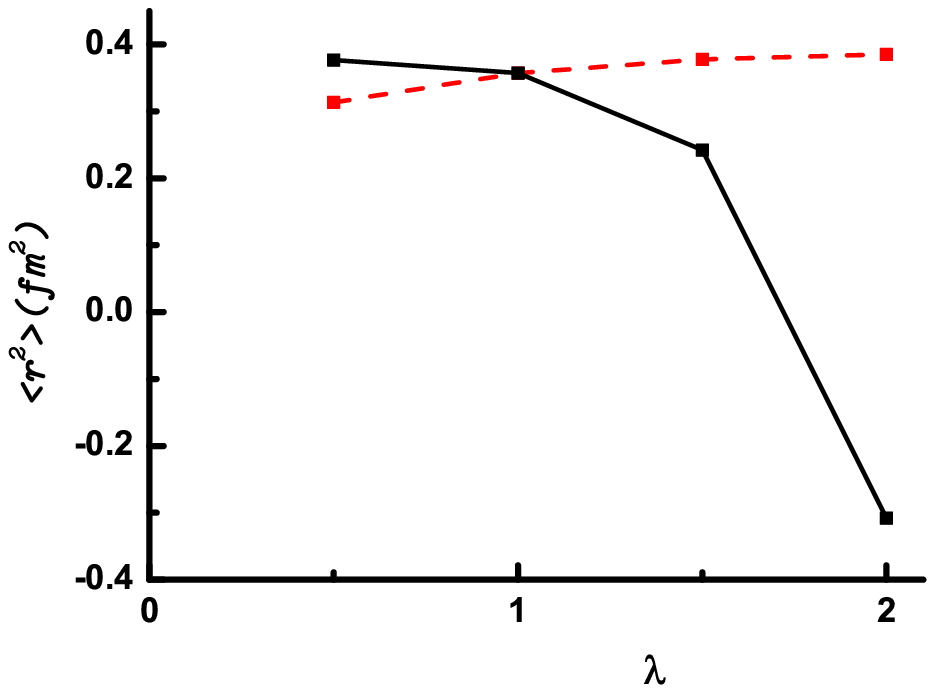}
\end{center}
\caption{Sensitivity of $\left\langle r^{2}\right\rangle $ in
$\operatorname{fm}^{2}$\ for the pion in relation with the propagator
parameters in the scenario S2.\ On the left we have $\left\langle
r^{2}\right\rangle $\ in relation with $\lambda\cdot\Lambda_{m}$ for
$\lambda=0.5,1,1.5$ and $2 $ (dashed curve) and $\left\langle r^{2}
\right\rangle $ in relation with $\lambda\cdot\alpha_{m}$ for the same values
of $\lambda$ (full curve). On the right, the dotted curve represents the
$\left\langle r^{2}\right\rangle $ in relation with $\lambda\cdot\Lambda_{z}$
\ and the full curve gives the $\left\langle r^{2}\right\rangle $\ for the
pion in relation with $\lambda\cdot\alpha_{z}$, for $\lambda=0.5,1,1.5$ and
$2.$ The full curve gives the msr in $\operatorname{fm}^{2}$\ for the pion in
relation with $\lambda\cdot\alpha_{z}$ for the same values of $\lambda.$}%
\label{Figrms}%
\end{figure}

We analyze the dependence of the pion radius in $m\left(  p\right)  $ and
$Z\left(  p\right)  $ in the chiral limit ($m_{0}=0).$ In Fig.~\ref{Figrms} we
show the result of rescaling by a generic factor of $\lambda$ the parameters
$\Lambda_{m}$ and $\alpha_{m}$ appearing in Eqs.(\ref{03.23mBowman}) and
(\ref{03.24ZBowman}). If $\Lambda_{m}$ increases the interaction $\vec{J}%
_{5}^{\,2}\left(  x\right)  $ becomes of shorter range. If we increase
$\alpha_{m},$ $g_{0}$ increases. In both cases the pion becomes more bound and
its radius smaller.

In Fig.~\ref{Figrms} we also show what happens to the mean square radius when
$\Lambda_{z}$ and $\alpha_{z}$ are rescaled by a factor of $\lambda$. The
system is not very sensitive to changes of $\Lambda_{z},$ while it is quite
sensitive for the rescaling of $\alpha_{z}$ when $\lambda\sim2.$ The reason
for this strong effect is that for this value $Z\left(  0\right)  =0.$

\section{Parton distribution.}

\label{SecPionPD}

\setcounter{equation}{0}

In the previous section we have analyzed some numerical aspects of the pion
form factor. We have put special emphasis in the discussion of the terms
restoring the local gauge symmetry, $\mathcal{V}_{1}\left(  p_{1}%
,p_{2}\right)  $ and $\mathcal{V}_{2}\left(  p_{1},p_{2}\right)  $. We have
seen that we have a second contribution given in equation (\ref{06.02FF4qf}),
but this contribution vanishes in our particular model.\ Searching for an
observable which is sensible to this term, we next discuss the parton
distribution. As it is shown in \cite{NogueraTheusslVento04} the operator for
the parton distribution can be connected to the electromagnetic operator at
zero momentum transfer. Therefore we are dealing with a property for
$k^{2}=0,$ where the path dependence is absent and our results will be valid
for any model.

The first step is to obtain the electromagnetic operator at zero momentum
transfer. This operator has a one body term which can be obtained directly
from Ward identity
\begin{equation}
\Gamma^{\mu}\left(  p,p\right)  =\frac{\partial S^{-1}\left(  p\right)
}{\partial p_{\mu}}\ , \label{07.00}%
\end{equation}
and a two body term which corresponds in our particular lagrangian to the
equation (\ref{06.02FF4qf}). To be precise, let us proceed to a general
discussion. We start from an action of the form
\begin{gather}
S=\int d^{4}x\bar{\psi}\left(  x\right)  \left(  i\not \partial -m_{0}\right)
\psi\left(  x\right)  +\nonumber\\
\int d^{4}x_{1}~d^{4}x_{2}~d^{4}x_{3}~d^{4}x_{4}\,G_{\alpha\beta\gamma\delta
}\left(  x_{1},x_{2},x_{3},x_{4}\right)  \bar{\psi}_{\delta}\left(
x_{4}\right)  \psi_{\beta}\left(  x_{2}\right)  \bar{\psi}_{\gamma}\left(
x_{3}\right)  \psi_{\alpha}\left(  x_{1}\right)  , \label{07.01}%
\end{gather}
where the greek indices characterize all symmetries, i.e spinor, color and
flavor. Let us define the following variables,
\begin{align}
x  &  =\left(  x_{3}-x_{1}\right)  \ ,\qquad x^{\prime}=\left(  x_{4}%
-x_{2}\right)  \ ,\nonumber\\
X^{\prime}  &  =\frac{1}{2}\left(  x_{3}+x_{1}-x_{4}-x_{2}\right)  \ ,\qquad
X=\frac{1}{4}\left(  x_{3}+x_{1}+x_{4}+x_{2}\right)  \ . \label{07.02}%
\end{align}
Translational invariance imposes that $G_{\alpha\beta\gamma\delta}\left(
x_{1},x_{2},x_{3},x_{4}\right)  $ cannot depend on $X.$ We introduce
\begin{equation}
G_{\alpha\beta\gamma\delta}\left(  x_{1},x_{2},x_{3},x_{4}\right)
=G_{\alpha\beta\gamma\delta}\left(  x,x^{\prime},X^{\prime}\right)  =\int
\frac{d^{4}p}{\left(  2\pi\right)  ^{4}}\frac{d^{4}p^{\prime}}{\left(
2\pi\right)  ^{4}}\frac{d^{4}P}{\left(  2\pi\right)  ^{4}}e^{-ixp}%
e^{-ix^{\prime}p^{\prime}}e^{-iX^{\prime}P}G_{\alpha\beta\gamma\delta}\left(
p,p^{\prime},P\right)  \ . \label{07.03}%
\end{equation}
Hermiticity and all internal symmetries, such as parity, charge conjugation
and time reversal, impose relations between the different components of the
interaction term $G_{\alpha\beta\gamma\delta}\left(  p,p^{\prime},P\right)  .$

We are interested in the mesonic bound state described in
Fig.~\ref{FigBetheSalpeter}. The Bethe-Salpeter amplitude for this meson is
defined by (we identify $p_{2}=p^{\prime}+P/2,$ $p_{4}=p^{\prime}-P/2,$
$p_{3}=p+P/2,$ $p_{1}=p-P/2)\ $
\begin{equation}
\Gamma_{\gamma\alpha}^{M}\left(  p,P\right)  =-2i\int\frac{d^{4}p^{\prime}%
}{\left(  2\pi\right)  ^{4}}G_{\alpha\beta\gamma\delta}\left(  p,p^{\prime
},P\right)  \left(  i\,S\left(  p^{\prime}+\frac{1}{2}P\right)  ~\Gamma
^{M}\left(  p^{\prime},P\right)  \,i\,S\left(  p^{\prime}-\frac{1}{2}P\right)
\right)  _{\beta\delta}\ \ , \label{07.05BSE}%
\end{equation}
with the standard normalization condition given in the Appendix
\ref{App4q-fGeneralVetrex}, equation (\ref{07.06.normalizacion}).

Regarding the charges of the fields in equation (\ref{07.01}), we can assume
that $Q_{\alpha}=Q_{\delta}=Q_{1}$ and $Q_{\gamma}=Q_{\beta}=Q_{2}.$ We
observe that the interaction term in equation (\ref{07.01}) is invariant under
global gauge transformations but not under local gauge transformations. One
way to make it locally invariant is by incorporating some links. The
interaction term becomes,
\begin{gather}
S_{i}=\int d^{4}x~d^{4}x^{\prime}~d^{4}X^{\prime}~d^{4}X\,G_{\alpha\beta
\gamma\delta}\left(  x,x^{\prime},X^{\prime}\right)  \bar{\psi}_{\delta
}\left(  \frac{1}{2}x^{\prime}-\frac{1}{2}X^{\prime}+X\right)  e^{-iQ_{1}%
\int_{X}^{\frac{1}{2}x^{\prime}-\frac{1}{2}X^{\prime}+X}dz^{\mu}A_{\mu}\left(
z\right)  }\nonumber\\
e^{-iQ_{2}\int_{-\frac{1}{2}x^{\prime}-\frac{1}{2}X^{\prime}+X}^{X}dz^{\mu
}A_{\mu}\left(  z\right)  }\psi_{\beta}\left(  -\frac{1}{2}x^{\prime}-\frac
{1}{2}X^{\prime}+X\right)  \bar{\psi}_{\gamma}\left(  \frac{1}{2}x+\frac{1}%
{2}X^{\prime}+X\right)  e^{-iQ_{2}\int_{X}^{\frac{1}{2}x+\frac{1}{2}X^{\prime
}+X}dz^{\mu}A_{\mu}\left(  z\right)  }\nonumber\\
e^{-iQ_{1}\int_{-\frac{1}{2}x+\frac{1}{2}X^{\prime}+X}^{X}dz^{\mu}A_{\mu
}\left(  z\right)  }\psi_{\alpha}\left(  -\frac{1}{2}x+\frac{1}{2}X^{\prime
}+X\right)  \ . \label{07.08}%
\end{gather}
We now expand this expression in powers of the photon field. The first term is
already included in equation (\ref{07.01}). The second term is linear in the
photon field and is the one of interest. We can evaluate it in the
$k\rightarrow0$ limit without having to define a specific path for the
integrals present in equation (\ref{07.08}). From this last result it is easy
to obtain that the quantity to add in the limit of $k\rightarrow0$ to each
vertex of the type of Fig.~\ref{Fig4quarkphoton} is
\begin{equation}
-iQ_{1}\left[  \Gamma_{1,\mu}^{(4q\gamma)}\left(  p_{1},p_{3};p_{2}%
,p_{4}\right)  \right]  _{\alpha\beta,\gamma\delta}-iQ_{2}\left[
\Gamma_{2,\mu}^{(4q\gamma)}\left(  p_{1},p_{3};p_{2},p_{4}\right)  \right]
_{\alpha\beta,\gamma\delta}\ , \label{07.11}%
\end{equation}
with
\begin{gather}
\left[  \Gamma_{1,\mu}^{(4q\gamma)}\left(  p_{1},p_{3};p_{2},p_{4}\right)
\right]  _{\alpha\beta,\gamma\delta}=2\left(  \frac{d}{dp_{1}^{\mu}}+\frac
{d}{dp_{4}^{\mu}}\right)  G_{\alpha\beta\gamma\delta}\left(  \frac{p_{1}%
+p_{3}}{2},\frac{p_{2}+p_{4}}{2},\frac{p_{2}+p_{3}-p_{1}-p_{4}}{2}\right)
~~,\nonumber\\
\left[  \Gamma_{2,\mu}^{(4q\gamma)}\left(  p_{1},p_{3};p_{2},p_{4}\right)
\right]  _{\alpha\beta,\gamma\delta}=2\left(  \frac{d}{dp_{2}^{\mu}}+\frac
{d}{dp_{3}^{\mu}}\right)  G_{\alpha\beta\gamma\delta}\left(  \frac{p_{1}%
+p_{3}}{2},\frac{p_{2}+p_{4}}{2},\frac{p_{2}+p_{3}-p_{1}-p_{4}}{2}\right)  ~~.
\label{07.12}%
\end{gather}
The details of the calculation are given in Appendix
\ref{App4q-fGeneralVetrex}, where we also discuss the implication of this term
to the form factor for a BSE with a $P-$dependent kernel.

In \cite{NogueraTheusslVento04}, the operator for the parton distribution has
been connected with the electromagnetic operator in the following way
\begin{gather}
q\left(  x\right)  =-\int\frac{d^{4}p}{\left(  2\pi\right)  ^{4}}%
\,\mathbb{T}\mathrm{r}\,\left[  iS\left(  p-\frac{1}{2}P\right)  \bar{\Gamma
}^{M}\left(  p,P\right)  i\,S\left(  p+\frac{1}{2}P\right)  \right.
\nonumber\\
\left.  \Gamma_{\mu}\left(  p+\frac{1}{2}P,p+\frac{1}{2}P\right)  n^{\mu
}i\,S\left(  p+\frac{1}{2}P\right)  \Gamma^{M}\left(  p,P\right)  \right]
\text{ }\delta\left(  x-\frac{n}{2}\cdot\left(  2p+P\right)  \right)  \ ,
\label{07.17}%
\end{gather}
where $\Gamma_{\mu}\left(  p,p\right)  $\ is the dressed photon vertex of the
selected parton, equation (\ref{07.00}). This expression can be generalized
from the one body coupling to include the two body coupling in the following
way
\begin{gather}
q\left(  x\right)  =\int\frac{d^{4}p}{\left(  2\pi\right)  ^{4}}\int
\frac{d^{4}p^{\prime}}{\left(  2\pi\right)  ^{4}}\,\left[  i\,S\left(
p^{\prime}-\frac{1}{2}P\right)  \bar{\Gamma}^{M}\left(  p,P\right)
i\,S\left(  p^{\prime}+\frac{1}{2}P\right)  \right]  _{\alpha\gamma
}\nonumber\\
\left[  \mathcal{O}_{u}\left(  p^{\prime}-\frac{1}{2}P,p^{\prime}+\frac{1}%
{2}P;p+\frac{1}{2}P,p-\frac{1}{2}P\right)  \right]  _{\alpha\beta,\gamma
\delta}\,\left[  \,i\,S\left(  p+\frac{1}{2}P\right)  \Gamma^{M}\left(
p,P\right)  i\,S\left(  p-\frac{1}{2}P\right)  \right]  _{\beta\delta}\ \ ,
\label{07.18}%
\end{gather}
with
\begin{gather}
\left[  \mathcal{O}_{u}\left(  p_{1},p_{3};p_{2},p_{4}\right)  \right]
_{\alpha\beta\gamma\delta}=-\left[  \Gamma_{\mu}\left(  p_{2},p_{3}\right)
n^{\mu}\right]  _{\gamma\beta}\text{ }\delta\left(  x-\frac{n}{2}\cdot\left(
p_{2}+p_{3}\right)  \right)  \left(  -i\right)  S_{\delta\alpha}^{-1}\left(
p_{1}\right)  \left(  2\pi\right)  ^{4}\delta^{4}\left(  p_{4}-p_{1}\right)
\nonumber\\
+\left[  \delta\left(  x-\frac{n}{2}\cdot\left(  p_{2}+P^{\prime}%
+p_{4}\right)  \right)  +\delta\left(  x-\frac{n}{2}\cdot\left(  p_{3}%
+P+p_{1}\right)  \right)  \right]  \ \frac{1}{2}n_{\mu}\left[  \Gamma_{2,\mu
}^{(4q\gamma)}\left(  p_{1},p_{3};p_{2},p_{4}\right)  \right]  _{\alpha
\beta,\gamma\delta}\ \ . \label{07.19}%
\end{gather}
\begin{figure}[ptb]
\begin{center}
\includegraphics[height=3.5366cm,width=8.9908cm]
{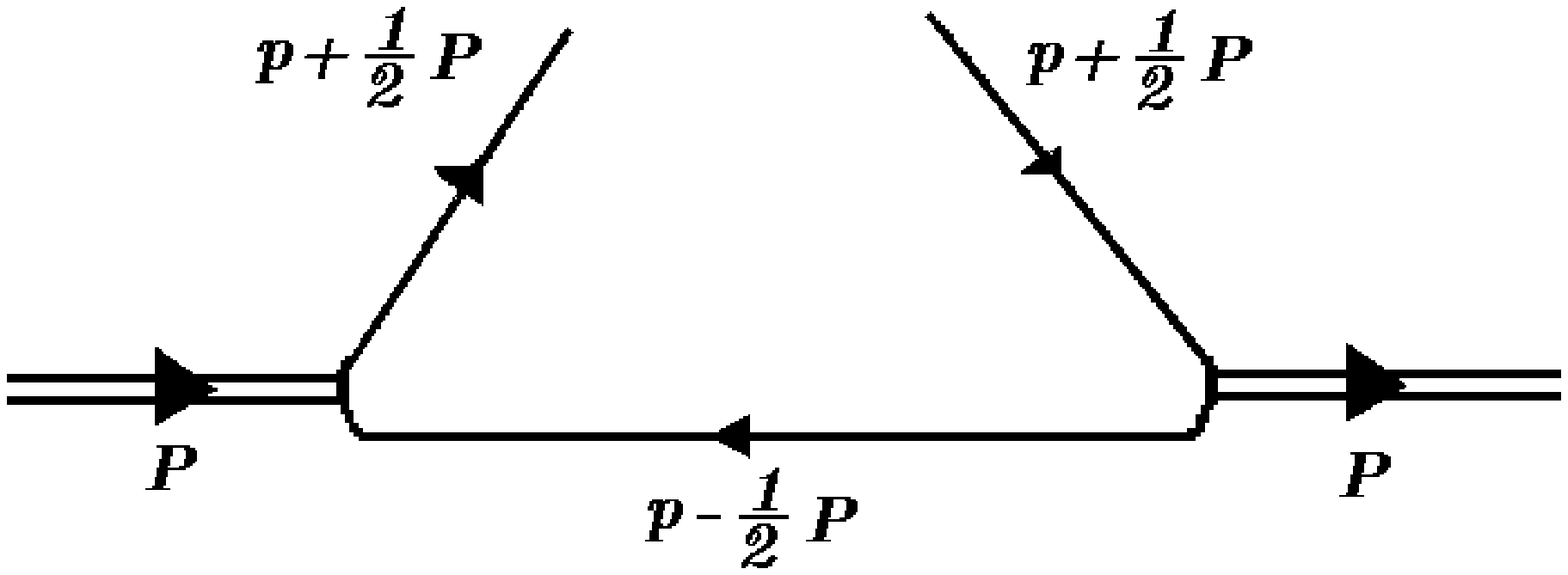}
\end{center}
\caption{Diagrammatic contributions to the parton distributions of the pion.}%
\label{FigPionPD1}%
\end{figure}\begin{figure}[ptbptb]
\begin{center}
\includegraphics[
trim=0.157471cm 0.000000cm 0.000000cm 0.000000cm,
height=3.0797cm,width=17.7883cm]
{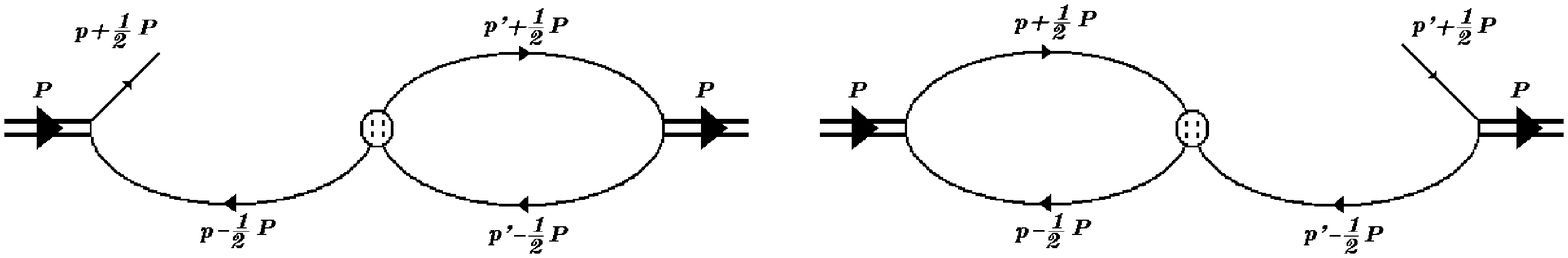}
\end{center}
\caption{Diagrammatic contributions to the parton distributions of the pion.}%
\label{FigPionPD2}%
\end{figure}The corresponding diagrams are those shown of
Figs.~\ref{FigPionPD1} and \ref{FigPionPD2}.

We can split the parton distribution into two one body and one two body
contributions
\begin{equation}
q\left(  x\right)  =q^{\left(  1\right)  }\left(  x\right)  +\tilde
{q}^{\left(  1\right)  }\left(  x\right)  +q^{\left(  2\right)  }\left(
x\right)  \ . \label{07.21}%
\end{equation}
The first contribution, $q^{\left(  1\right)  }\left(  x\right)  ,$ is the one
appearing in equation (\ref{07.17}) and corresponds to the diagram shown in
Fig.~\ref{FigPionPD1}. The second term is also a one body term,
\begin{gather}
\tilde{q}^{\left(  1\right)  }\left(  x\right)  =\int\frac{d^{4}p}{\left(
2\pi\right)  ^{4}}\frac{1}{4}\delta\left(  x-\frac{n}{2}\cdot\left(
2p+P\right)  \right)  \mathbb{T}\mathrm{r}\left[  in_{\mu}~\frac{d\bar{\Gamma
}^{M}\left(  p,P\right)  }{dp_{\mu}}\,i\,S\left(  p+\frac{1}{2}P\right)
\Gamma^{M}\left(  p,P\right)  i\,S\left(  p-\frac{1}{2}P\right)  \right.
\nonumber\\
+\left.  i\,S\left(  p-\frac{1}{2}P\right)  \bar{\Gamma}^{M}\left(
p,P\right)  i\,S\left(  p+\frac{1}{2}P\right)  in_{\mu}~\frac{d\Gamma
^{M}\left(  p,P\right)  }{dp_{\mu}}\right]  \ , \label{07.22}%
\end{gather}
while the third term is a genuine two body term given by,
\begin{gather}
q^{\left(  2\right)  }\left(  x\right)  =\int\frac{d^{4}p}{\left(
2\pi\right)  ^{4}}\int\frac{d^{4}p^{\prime}}{\left(  2\pi\right)  ^{4}%
}\,\left[  i\,S\left(  p^{\prime}-\frac{1}{2}P\right)  \bar{\Gamma}^{M}\left(
p^{\prime},P\right)  i\,S\left(  p^{\prime}+\frac{1}{2}P\right)  \right]
_{\alpha\gamma}\nonumber\\
\frac{1}{2}n_{\mu}~\left\{  \left[  \delta\left(  x-\frac{n}{2}\cdot\left(
2p+P\right)  \right)  \left(  \frac{d}{dp_{\mu}^{\prime}}+2\frac{d}{dP_{\mu}%
}\right)  +\delta\left(  x-\frac{n}{2}\cdot\left(  2p^{\prime}+P\right)
\right)  \left(  \frac{d}{dp_{\mu}}+2\frac{d}{dP_{\mu}}\right)  \right]
G_{\alpha\beta\gamma\delta}\left(  p^{\prime},p,P\right)  \right\} \nonumber\\
\left[  \,i\,S\left(  p+\frac{1}{2}P\right)  \Gamma^{M}\left(  p,P\right)
i\,S\left(  p-\frac{1}{2}P\right)  \right]  _{\beta\delta}\ . \label{07.23}%
\end{gather}
These last two contributions correspond to the diagram of Fig~\ref{FigPionPD2}
. In $\tilde{q}^{\left(  1\right)  }\left(  x\right)  $ the bubble integral
has been performed using the BSE, that is the reason for appearing as a one
body term.

The whole contribution coming from the diagram of Fig~\ref{FigPionPD2} has a
non-vanishing contributions to $q\left(  x\right)  ,$ nevertheless, when we
integrate over $x$ the only non-vanishing contribution that can survive is the
one associated to the derivative of the total momentum, which contributes to
the form factor. An analysis of this operator in the scenarios here considered
is done in \cite{NogueraVento05}.

Summarizing, we have three contributions: the standard one, associated with
the handbag diagram, and defined in equation (\ref{07.17}) and two new
contributions defined in Eqs. (\ref{07.22}) and (\ref{07.23}). These
contributions are a consequence of the non locality of the currents involved
in our model. From the point of view of QCD they are also handbag diagrams,
but in terms of the BS amplitudes they have a different structure.

\section{Conclusion.}

We have defined a family of phenomenological chirally invariant non local
lagrangians to describe hadron structure. They lead to non trivial momentum
dependencies in the quark propagator parametrized by momentum dependent quark
mass terms and wave function renormalization constants. We have shown that the
formalism is able to emulate any propagator obtained from more fundamental
studies. In particular we have studied two scenarios obtained from low energy
models of QCD \cite{DyakonovPetrov86}, and lattice QCD \cite{Bowman02,
Bowman03}. As a first goal, our lagrangian description allows a careful study
of the properties of any observable. In particular we have applied it\ in
detail to the study of the pion electromagnetic form factor with special
emphasis to the consequences of the local gauge invariance.

Several authors have previously used non local models, in particular the non
local generalization of the Nambu- Jona Lasinio model \cite{Birse95,
Birse98,Osipov95, GomezDummScoccola04,Scoccola01-04} and the instanton liquid
model \cite{Dorokhov1}, and have studied electromagnetic properties
\cite{Dorokhov2, Dorokhov3}. There are main differences between these
approaches and ours. The momentum dependence in our case arises from QCD and
appears not only in the mass but also in the wave function renormalization.
Moreover, we do not use a separable approximation of the interaction kernel in
each particle momentum.

In order to test the consistency of the model, we have studied the basic
properties of the theory, i. e. effective current masses and quark condensate.
We have constructed, using our models, the two body quark anti-quark bound
state equation and solved for the pion obtaining its mass and its
Bethe-Salpeter amplitude.

We have implemented the electromagnetic coupling in this theory. The
construction of the dressed quark-photon vertex is complicated and the WTI do
not solve the problem completely. In particular the transverse propagator has
been a subject of much debate. In our lagrangian formalism it is natural to
implement local gauge invariance through links between the points where the
quark fields act. We find that two new terms appear in the quark-photon vertex
which restore local gauge symmetry. We have obtained simple expressions for
these two new terms choosing the link between the two points characterizing
the non local currents as a straight line. They appear as derivatives of the
mass, $m\left(  p\right)  $, and wave function renormalization, $Z\left(
p\right)  $, of the quarks. We apply our formalism to the pion form factor and
find out that the local gauge restoring terms could amount to as much as
20-30\%. Our analysis is applicable to previous work, where the Ball-Chiu
prescription for the electromagnetic vertex were used.

We have applied the same ideas to the axial current, implementing local gauge
invariance under SU$_{L}\left(  2\right)  $ transformations in the lagrangian.
We have obtained the full dressed axial vertex. We have discussed several
equivalent ways for calculating the pion decay constant. The calculated pion
decay constant is in reasonably good agreement with the experimental result.
Moreover, we can observe that the two scenarios studied in the paper describe
the same physics. It must be emphasized that this two scenarios have a quite
different origin.

The basic relations from chiral symmetry, the Goldberger-Treiman and the
Gell-Mann-Oakes-Renner relations, have been recovered.

The simplest expression for the axial current, given in equation
(\ref{05a.31}), is a good approximation for the calculation of $f_{\pi}$.
Nevertheless it does not include the pion pole contribution.\ Equation
(\ref{05a.33}) provides an expression for the longitudinal part of the axial
current consistent with PCAC.

Another effect of local gauge invariance is that a new vertex with four quark
lines and one photon line appears. The contribution of this vertex to the pion
form factor vanishes for our models. This is not a general result, but a
consequence of the separable nature and $P-$independence of the interacting
terms in our lagrangian. For a general kernel we have seen that this term is
necessary in order to guarantee the normalization of the form factor,
$F\left(  0\right)  =1.$

Searching for an observable which is sensitive to this term we have studied
the parton distribution. Following the ideas of ref
\cite{NogueraTheusslVento04} we have obtained the related operator which is
path independent and therefore can be used in any other model. The new two
body term will give a non vanishing contribution to the parton distribution
even within our models \cite{NogueraVento05}.

Regarding the numerical values we observe that the two studied scenarios give
quite similar results. We only fit $m_{\mathbf{0}}$\ in order to reproduce
$m_{\pi}$. Their analytic form for the mass, given in equation
(\ref{03.22DyakonovPetrov0}) and (\ref{03.23mBowman}), are quite different,
but their numerical value in the region $p_{E}^{2}<6\operatorname{GeV}^{2}$ is
similar. We regard these expressions as approximations of more realistic
expressions for the propagator. In this way, their unwanted analytic
properties are not considered. We conclude that both scenarios provide an
overall good agreement with the data. The value of the condensate is well
reproduced, the value of $m_{0}$ is in agreement with the quark current mass,
the value of the pion decay constant is 10\% smaller than the experimental
value, and both scenarios give similar results. We observe that the fact that
$Z\left(  p\right)  \neq1$\ in the second scenario is not important in these
observables. The mean square radius discriminates between these scenarios. In
the S1 (S2) we obtain a value which is 10\% (20\%) smaller than the
experimental one. Our analysis shows that this difference is not associated to
the different choice of the mass expression but to the fact that in the S2
$Z\left(  p\right)  \neq1.$ The electromagnetic radius of the pion will be
affected by the coupling of the photon to the meson vectors. This contribution
has been estimated to be around the 10-20\% \cite{MarisTandi00,
OconnellPearceThomasWilliams97}. The intermediate region, $k^{2}%
\sim0.6-0.8\operatorname{GeV}^{2}$, will be also affected by these vector
currents through axial components in the pion Bethe-Salpeter amplitude.
Therefore we cannot conclude which of the two models is more accurate at present.

In summary we have set the formalism for the description of non local models
of hadron structure and have used them to analyze past developments and
propose future studies. In particular, it has been already applied for the
study of the parton distribution of the pion \cite{NogueraVento05}.

\begin{acknowledgments}
I wish to thank Prof. V. Vento for many useful discussions. This work was
supported by the sixth framework programme of the European Commission under
contract 506078 (I3 Hadron Physics), MEC (Spain) under contract
FPA2004-05616-C02-01, and Generalitat Valenciana under contract GRUPOS03/094.
\end{acknowledgments}

\appendix{}

\section{The quark photon vertex}

\label{AppQuarkPhotonVertex}

Let us detail here some intermediate steps related with section
\ref{SecQuarkPhotonVertex}.

The set of eight basic transverse tensors $T_{i}^{\mu}\left(  p_{1}%
,p_{2}\right)  $ we use in equation (\ref{05.04quarkphotonvertex}) are those
defined by Ball and Chiu \cite{BallChiu80} with some minor modifications
$\left(  k=p_{2}-p_{1}\right)  $,
\begin{align}
T_{1}^{\mu}\left(  p_{1},p_{2}\right)   &  =p_{1}^{\mu}\left(  p_{2}.k\right)
-p_{2}^{\mu}\left(  p_{1}.k\right)  \;,\nonumber\\
T_{2}^{\mu}\left(  p_{1},p_{2}\right)   &  =\left[  p_{1}^{\mu}\left(
p_{2}.k\right)  -p_{2}^{\mu}\left(  p_{1}.k\right)  \right]  \frac{\left(
\rlap{$/$}p_{1}+\rlap{$/$}p_{2}\right)  }{2}\;,\nonumber\\
T_{3}^{\mu}\left(  p_{1},p_{2}\right)   &  =k^{2}\gamma^{\mu}-k^{\mu
}\rlap{$/$}k\;,\nonumber\\
T_{4}^{\mu}\left(  p_{1},p_{2}\right)   &  =-i\left[  p_{1}^{\mu}\left(
p_{2}.k\right)  -p_{2}^{\mu}\left(  p_{1}.k\right)  \right]  p_{1}^{\lambda
}p_{2}^{\nu}\sigma_{\lambda\nu}\;,\nonumber\\
T_{5}^{\mu}\left(  p_{1},p_{2}\right)   &  =-i\sigma^{\mu\lambda}k_{\lambda
}\;,\label{0A.01}\\
T_{6}^{\mu}\left(  p_{1},p_{2}\right)   &  =\gamma^{\mu}\left(  p_{2}%
^{2}-p_{1}^{2}\right)  -\left(  p_{1}+p_{2}\right)  ^{\mu}%
\rlap{$/$}k\;,\nonumber\\
T_{7}^{\mu}\left(  p_{1},p_{2}\right)   &  =\frac{p_{2}^{2}-p_{1}^{2}}%
{2}\left[  \gamma^{\mu}\left(  \rlap{$/$}p_{1}+\rlap{$/$}p_{2}\right)
-p_{1}^{\mu}-p_{2}^{\mu}\right]  -i\left(  p_{1}+p_{2}\right)  ^{\mu}%
p_{1}^{\lambda}p_{2}^{\nu}\sigma_{\lambda\nu}\;,\nonumber\\
T_{8}^{\mu}\left(  p_{1},p_{2}\right)   &  =i\gamma^{\mu}p_{1}^{\lambda}%
p_{2}^{\nu}\sigma_{\lambda\nu}+p_{1}^{\mu}\rlap{$/$}p_{2}-p_{2}^{\mu
}\rlap{$/$}p_{1}\;.\nonumber
\end{align}

\bigskip

Looking at the four quark photon vertex, we must expand the exponential with
the $A_{\mu}\left(  z\right)  $ field in the currents of
Eqs.(\ref{05.07aScalarGaugeICurrent}-\ref{05.07cMomentumGaugeICurrent}). Let
us start with the scalar current equation (\ref{05.07aScalarGaugeICurrent})
\begin{subequations}
\label{0A.06Scalar1A}%
\begin{align}
J_{S}\left(  x\right)   &  =J_{S}^{\left(  0A\right)  }\left(  x\right)
+J_{S}^{\left(  1A\right)  }\left(  x\right)  +....=\int d^{4}y\,G_{0}\left(
y\right)  \bar{\psi}\left(  x+\frac{1}{2}y\right)  ~\psi\left(  x-\frac{1}%
{2}y\right)  +\label{0A.06aScalar0}\\[0.75cm]
&  \int d^{4}y\,G_{0}\left(  y\right)  \bar{\psi}\left(  x+\frac{1}%
{2}y\right)  ~\left(  -iQ\int_{x-\frac{1}{2}y}^{x+\frac{1}{2}y}dz^{\mu}A_{\mu
}\left(  z\right)  \right)  ~\psi\left(  x-\frac{1}{2}y\right)  +...\ \ \ .
\label{0A.06bScalar1A}%
\end{align}%
\end{subequations}
\[
\]
Inserting this expansion in the $J_{S}^{\dagger}\left(  x\right)  J_{S}\left(
x\right)  $ term of equation (\ref{02.04Lagrangian}), the first crossed term,
$J_{S}^{\left(  0A\right)  \dag}\left(  x\right)  J_{S}^{\left(  1A\right)
}\left(  x\right)  +J_{S}^{\left(  1A\right)  \dag}\left(  x\right)
J_{S}^{\left(  0A\right)  }\left(  x\right)  ,$ leads to the 4 quarks-photon
vertex. An evaluation of this vertex needs to specify the path from
$x-\frac{1}{2}y$ to $x+\frac{1}{2}y$ followed by the evaluation of the
integral on $z^{\mu}$ in equation (\ref{0A.06bScalar1A}). This path can be
parameterized as
\begin{equation}
z^{\mu}=x^{\mu}+\frac{\lambda}{2}y^{\mu}+c_{\text{\textsc{S}}}^{\mu}\left(
\lambda\right)  \text{~~~~with~~~~}\lambda\in\left[  -1,1\right]
\text{~~~~and~~~~}c_{\text{\textsc{S}}}^{\mu}\left(  -1\right)
=c_{\text{\textsc{S}}}^{\mu}\left(  1\right)  =0\ \ .
\label{0A.07caminoScalar}%
\end{equation}

The current $J_{S}^{\left(  1A\right)  }\left(  x\right)  $ between states of
1 quark of momentum $p_{1},$ and 1 photon-1 quark of momentum $k$ and $p_{3},$
respectively gives
\begin{align}
\left\langle p_{3}\left\vert J_{S}^{\left(  1A\right)  }\left(  x\right)
\right\vert p_{1}k\right\rangle  &  =\bar{u}\left(  p_{3}\right)  \left(
-iQ\right)  u\left(  p_{1}\right)  e^{i\left(  p_{3}-k-p_{1}\right)
x}\epsilon^{\mu}\left(  k,\xi\right)  \int\frac{d^{4}t}{\left(  2\pi\right)
^{4}}G_{0}\left(  t\right) \nonumber\\
&  \int d^{4}y\,\left(  \int_{-1}^{1}d\lambda\left(  \frac{1}{2}y_{\mu
}+c_{\text{\textsc{S}}\mu}^{\prime}\right)  e^{-ikc_{\text{\textsc{S}}}%
}\right)  e^{iy\left(  \frac{p_{1}+p_{3}-\lambda k}{2}-t\right)  }\nonumber\\
&  =\bar{u}\left(  p_{3}\right)  \left(  -iQ\right)  u\left(  p_{1}\right)
e^{i\left(  p_{3}-k-p_{1}\right)  x}\epsilon^{\mu}\left(  k,\xi\right)
\nonumber\\
&  \int_{-1}^{1}d\lambda\;e^{ikc_{\text{\textsc{S}}}}\;\left(  -\left(
\frac{p_{1}+p_{3}-\lambda k}{2}\right)  _{\mu}iG_{0}^{\prime}\left(  t\right)
+c_{\text{\textsc{S}}\mu}^{\prime}G_{0}\left(  t\right)  \right)
_{t=\frac{p_{1}+p_{3}-\lambda k}{2}}\ , \label{0A.08}%
\end{align}
where $Q=e\left(  \vec{\tau}.\hat{n}+1/3\right)  /2$ is the charge of the
quark, $G_{0}^{\prime}\left(  t\right)  =dG_{0}\left(  t\right)  /dt^{2}$ and
$\epsilon^{\mu}\left(  k,\xi\right)  $ is the photon polarization. For the
sake of simplicity we take $c_{\text{\textsc{S}}\mu}=0$ which corresponds to
linking the two points of the non local current by a straight line. We can
write the matrix element in the following way
\begin{gather}
\left\langle p_{3}\left\vert J_{S}^{\left(  1A\right)  }\left(  x\right)
\right\vert p_{1}k\right\rangle =-\bar{u}\left(  p_{3}\right)  ~Q~u\left(
p_{1}\right)  e^{i\left(  p_{3}-k-p_{1}\right)  x}\epsilon^{\mu}\left(
k,\xi\right) \nonumber\\
\left[  \left(  p_{1}+p_{3}\right)  _{\mu}\mathbb{V}_{0a}\left(  \frac
{p_{1}+p_{3}}{2},k\right)  -k_{\mu}\mathbb{V}_{0b}\left(  \frac{p_{1}+p_{3}%
}{2},k\right)  \right]  \ , \label{0A.10}%
\end{gather}
where
\begin{subequations}
\label{0A.11}%
\begin{gather}
\mathbb{V}_{0a}\left(  \bar{p},k\right)  =\int_{-1}^{1}d\lambda\;\frac{1}%
{2}G_{0}^{\prime}\left(  \bar{p}-\lambda\frac{k}{2}\right)  \ , \label{0A.11a}%
\\[0.5cm]
\mathbb{V}_{0b}\left(  \bar{p},k\right)  =\int_{-1}^{1}d\lambda\;\frac
{\lambda}{2}\;G_{0}^{\prime}\left(  \bar{p}-\lambda\frac{k}{2}\right)  \ .
\label{0A.11b}%
\end{gather}
Using the current given in equation (\ref{0A.10}) we obtain the vertex
associated to the interacting term $g_{0}\left(  J_{S}^{\left(  0A\right)
\dag}\left(  x\right)  J_{S}^{\left(  1A\right)  }\left(  x\right)
+J_{S}^{\left(  1A\right)  \dag}\left(  x\right)  J_{S}^{\left(  0A\right)
}\left(  x\right)  \right)  $ of the Lagrangian,
\end{subequations}
\begin{gather}
\left[  \Gamma_{S,\mu}^{(4q\gamma)}\left(  p_{1},p_{3};p_{2},p_{4}\right)
\right]  _{\alpha\beta,\gamma\delta}=-i2g_{0}G_{0}\left(  \frac{p_{2}+p_{4}%
}{2}\right)  \delta_{\beta\delta}Q_{\gamma\alpha}\left[  \left(  p_{1}%
+p_{3}\right)  _{\mu}\mathbb{V}_{0a}\left(  \frac{p_{1}+p_{3}}{2},k\right)
\right. \nonumber\\
-\left.  k_{\mu}\mathbb{V}_{0b}\left(  \frac{p_{1}+p_{3}}{2},k\right)
\right]  +\left(  p_{1}\alpha,p_{3}\gamma\leftrightarrow p_{2}\beta
,p_{4}\delta\right)  \ , \label{0A.12Scalar4q-fVertex}%
\end{gather}
with $k=p_{3}+p_{4}-p_{1}-p_{2}.$

\bigskip

We apply the same ideas to the pseudoscalar current
(\ref{05.07bPseudoscalarGaugeICurrent}) obtaining the vertex associated to the
interacting term $g_{0}\left(  J_{5}^{\left(  0A\right)  \dag}\left(
x\right)  J_{5}^{\left(  1A\right)  }\left(  x\right)  +J_{5}^{\left(
1A\right)  \dag}\left(  x\right)  J_{5}^{\left(  0A\right)  }\left(  x\right)
\right)  $ of the Lagrangian,
\begin{align}
&  \left[  \Gamma_{5,\mu}^{(4q\gamma)}\left(  p_{1},p_{3};p_{2},p_{4}\right)
\right]  _{\alpha\beta,\gamma\delta}=-i2g_{0}G_{0}\left(  \frac{p_{2}+p_{4}%
}{2}\right)  \left(  i\gamma_{5}\vec{\tau}\right)  _{\delta\beta}\nonumber\\
&  \left\{  \left(  i\gamma_{5}\frac{1}{2}\left\{  Q,\vec{\tau}\right\}
\right)  _{\gamma\alpha}\left[  \left(  p_{1}+p_{3}\right)  _{\mu}%
\mathbb{V}_{0a}\left(  \frac{p_{1}+p_{3}}{2},k\right)  -k_{\mu}\mathbb{V}%
_{0b}\left(  \frac{p_{1}+p_{3}}{2},k\right)  \right]  \right.  +\nonumber\\
&  \left.  \left(  i\gamma_{5}\frac{1}{2}\left[  Q,\vec{\tau}\right]  \right)
_{\gamma\alpha}\left[  \left(  p_{1}+p_{3}\right)  _{\mu}\mathbb{A}%
_{0a}\left(  \frac{p_{1}+p_{3}}{2},k\right)  -k_{\mu}\mathbb{A}_{0b}\left(
\frac{p_{1}+p_{3}}{2},k\right)  \right]  \right\}  +\left(  p_{1}\alpha
,p_{3}\gamma\leftrightarrow p_{2}\beta,p_{4}\delta\right)  \ ,
\label{0A.17Pseudo4q-fVertex}%
\end{align}
with
\begin{subequations}
\label{0A.16}%
\begin{align}
\mathbb{A}_{0a}\left(  \bar{p},k\right)   &  =\int_{0}^{1}d\lambda\;\frac
{1}{2}G_{0}^{\prime}\left(  \bar{p}-\lambda\frac{k}{2}\right)  -\int_{-1}%
^{0}d\lambda\;\frac{1}{2}G_{0}^{\prime}\left(  \bar{p}-\lambda\frac{k}%
{2}\right)  \ ,\label{0A.16c}\\[0.5cm]
\mathbb{A}_{0b}\left(  \bar{p},k\right)   &  =\int_{0}^{1}d\lambda
\;\frac{\lambda}{2}\;G_{0}^{\prime}\left(  \bar{p}-\lambda\frac{k}{2}\right)
-\int_{-1}^{0}d\lambda\;\frac{\lambda}{2}\;G_{0}^{\prime}\left(  \bar
{p}-\lambda\frac{k}{2}\right)  \ . \label{0A.16d}%
\end{align}
and $\mathbb{V}_{0a}$ and $\mathbb{V}_{0b}$ given in equations (\ref{0A.11a})
and (\ref{0A.11b}).

\bigskip

The momentum current equation (\ref{05.07cMomentumGaugeICurrent}) produces the
interacting term $g_{p}\left(  J_{p}^{\left(  0A\right)  \dag}\left(
x\right)  J_{p}^{\left(  1A\right)  }\left(  x\right)  +J_{p}^{\left(
1A\right)  \dag}\left(  x\right)  J_{p}^{\left(  0A\right)  }\left(  x\right)
\right)  .$ The associated vertex is%
\end{subequations}
\begin{gather}
\left[  \Gamma_{p,\mu}^{(4q\gamma)}\left(  p_{1},p_{3};p_{2},p_{4}\right)
\right]  _{\alpha\beta,\gamma\delta}=-i2g_{p}G_{p}\left(  \frac{p_{2}+p_{4}%
}{2}\right)  \left(  \frac{\rlap{$/$}p_{2}+\rlap{$/$}p_{4}}{2}\right)
_{\delta\beta}\left(  Q\gamma^{\nu}\right)  _{\gamma\alpha}\nonumber\\[0.1in]
\left\{  g_{\nu\mu}\frac{1}{2}\left[  G_{p}\left(  \frac{p_{1}+p_{3}+k}%
{2}\right)  +G_{p}\left(  \frac{p_{1}+p_{3}-k}{2}\right)  \right]  \right.
\nonumber\\
+\left.  \frac{\left(  p_{1}+p_{3}\right)  _{\nu}}{2}\left[  \left(
p_{1}+p_{3}\right)  _{\mu}\mathbb{V}_{pa}\left(  \frac{p_{1}+p_{3}}%
{2},k\right)  -k_{\mu}\mathbb{V}_{pb}\left(  \frac{p_{1}+p_{3}}{2},k\right)
\right]  \right\}  +\left(  p_{1}\alpha,p_{3}\gamma\leftrightarrow p_{2}%
\beta,p_{4}\delta\right)  \ , \label{0A.22Momento4q-fVertex}%
\end{gather}
whith
\begin{subequations}
\label{0A.21}%
\begin{gather}
\mathbb{V}_{pa}\left(  \bar{p},k\right)  =\int_{-1}^{1}d\lambda\;\frac{1}%
{2}~G_{p}^{\prime}\left(  \bar{p}-\lambda\frac{k}{2}\right)  \ ,
\label{0A.21a}\\[0.5cm]
\mathbb{V}_{pb}\left(  \bar{p},k\right)  =\int_{-1}^{1}d\lambda\;\frac
{\lambda}{2}\;G_{p}^{\prime}\left(  \bar{p}-\lambda\frac{k}{2}\right)  \ .
\label{0A.21b}%
\end{gather}%
\end{subequations}
\[
\]

\section{The quark $W^{\mu}$ vertex.}

\label{AppQuarkWVertex}

Let us decompose the quark field in its left a right parts, $\psi^{L,R}\left(
x\right)  =\frac{1}{2}\left(  1\mp\gamma_{5}\right)  \psi\left(  x\right)  .$
Under a local $SU_{L}\left(  2\right)  $ gauge transformations we have%
\begin{align}
\psi^{L}\left(  x\right)   &  \longrightarrow e^{ig_{w}\frac{\vec{\tau}}%
{2}.\vec{\alpha}\left(  x\right)  }\psi^{L}\left(  x\right)  \qquad\psi
^{R}\left(  x\right)  \longrightarrow\psi^{R}\left(  x\right) \label{0B.02}\\
\vec{\tau}\vec{W}^{\mu}\left(  x\right)   &  =\vec{\tau}\vec{W}^{\mu}\left(
x\right)  -\partial^{\mu}\vec{\alpha}\left(  x\right)  .\vec{\tau}%
-g_{w}~\left(  \vec{\alpha}\left(  x\right)  \times\vec{W}^{\mu}\left(
x\right)  \right)  .\vec{\tau}~~. \label{0B.03}%
\end{align}
Our lagrangian model defined in (\ref{02.04Lagrangian}) is invariant under
infinitesimal global gauge $SU_{L}\left(  2\right)  $ transformation. Now we
are interested in making this lagrangian invariant under local
transformations. We proceed in a similar way as in the electromagnetic case.
The main difficulty is on the fact that our currents are built with fields in
different points, thus the value of $\vec{\alpha}\left(  x\right)  $ will be
different for each point. To avoid this difficulty let us introduce the
following product of fields%
\begin{equation}
\mathcal{P}\left(  e^{-i\frac{g_{w}}{2}\int_{y}^{x}dz^{\mu}~\vec{\tau}\vec
{W}^{\mu}\left(  z\right)  }\right)  ~\psi^{L}\left(  y\right)  \label{0B.04}%
\end{equation}
which transform covariantly at least for infinitesimal transformations,%
\begin{equation}
\left[  e^{-i\frac{g_{w}}{2}\int_{y}^{x}dz^{\mu}~\vec{\tau}\vec{W}^{\mu
}\left(  z\right)  }~\psi^{L}\left(  y\right)  \right]  \longrightarrow
e^{i\frac{g_{w}}{2}\vec{\tau}.\vec{\alpha}\left(  x\right)  }\left[
e^{-i\frac{g_{w}}{2}\int_{y}^{x}dz^{\mu}~\vec{\tau}\vec{W}^{\mu}\left(
z\right)  +\mathcal{O}\left(  g_{w}^{2}\right)  }~\psi^{L}\left(  y\right)
\right]  \label{0B.05}%
\end{equation}
Then, we can define the currents%
\begin{align}
J_{S}\left(  x\right)   &  =\int d^{4}y\,G_{0}\left(  y\right)  \left[
\bar{\psi}^{R}\left(  x+\frac{1}{2}y\right)  ~\mathcal{P}\left(
e^{-i\frac{g_{w}}{2}\int_{x-\frac{1}{2}y}^{x}dz^{\mu}\vec{W}_{\mu}\left(
z\right)  \vec{\tau}}\right)  ~\psi^{L}\left(  x-\frac{1}{2}y\right)  \right.
\nonumber\\[0.3in]
&  ~~~~~~~~~~~~~~~~~~~~+\left.  \bar{\psi}^{L}\left(  x+\frac{1}{2}y\right)
~\mathcal{P}\left(  e^{-i\frac{g_{w}}{2}\int_{x}^{x+\frac{1}{2}y}dz^{\mu}%
\vec{W}_{\mu}\left(  z\right)  \vec{\tau}}\right)  ~\psi^{R}\left(  x-\frac
{1}{2}y\right)  \right]  \ , \label{0B.06}%
\end{align}%
\begin{align}
\vec{J}_{5}\left(  x\right)   &  =\int d^{4}y\,G_{0}\left(  y\right)  \left[
-i~\bar{\psi}^{R}\left(  x+\frac{1}{2}y\right)  ~\vec{\tau}~\mathcal{P}\left(
e^{-i\frac{g_{w}}{2}\int_{x-\frac{1}{2}y}^{x}dz^{\mu}\vec{W}_{\mu}\left(
z\right)  \vec{\tau}}\right)  ~\psi^{L}\left(  x-\frac{1}{2}y\right)  \right.
\nonumber\\[0.3in]
&  ~~~~~~~~~~~~~~~~~~~~+\left.  i~\bar{\psi}^{L}\left(  x+\frac{1}{2}y\right)
~\mathcal{P}\left(  e^{-i\frac{g_{w}}{2}\int_{x}^{x+\frac{1}{2}y}dz^{\mu}%
\vec{W}_{\mu}\left(  z\right)  \vec{\tau}}\right)  ~\vec{\tau}~\psi^{R}\left(
x-\frac{1}{2}y\right)  \right]  \ , \label{0B.07}%
\end{align}%
\begin{align}
J_{p}\left(  x\right)   &  =\int d^{4}y\,G_{p}\left(  y\right)  \frac{1}%
{2}\left[  \bar{\psi}^{L}\left(  x+\frac{1}{2}y\right)  ~\mathcal{P}\left(
e^{-i\frac{g_{w}}{2}\int_{x-\frac{1}{2}y}^{x+\frac{1}{2}y}dz^{\mu}\vec{W}%
_{\mu}\left(  z\right)  \vec{\tau}}\right)  ~i\ \rlap{$/$}\hspace
{-0.05cm}D\ \psi^{L}\left(  x-\frac{1}{2}y\right)  \right. \nonumber\\
&  ~~~~~~~~~~~~~~~~\ \ \ \ \ \ \ \ -i\bar{\psi}^{L}\left(  x+\frac{1}%
{2}y\right)  \overleftarrow{\rlap{$/$}\hspace*{-0.05cm}D}~\mathcal{P}\left(
e^{-i\frac{g_{w}}{2}\int_{x-\frac{1}{2}y}^{x+\frac{1}{2}y}dz^{\mu}\vec{W}%
_{\mu}\left(  z\right)  \vec{\tau}}\right)  ~\psi^{L}\left(  x-\frac{1}%
{2}y\right) \nonumber\\
&  ~~~~~~~~~~~~~~~~~~~~+\left.  i\bar{\psi}^{R}\left(  x+\frac{1}{2}y\right)
\overleftrightarrow{\rlap{$/$}\hspace*{-0.05cm}\partial}~\psi^{R}\left(
x-\frac{1}{2}y\right)  \right]  \ , \label{0B.08}%
\end{align}
where the covariant derivative is $D^{\mu}\psi\left(  x\right)  =\left[
\partial^{\mu}+ig_{w}\vec{W}^{\mu}\left(  z\right)  \frac{\vec{\tau}}%
{2}\right]  \psi^{L}\left(  x\right)  .$ With these definitions, the
combination $J_{S}^{2}\left(  x\right)  +\vec{J}_{5}^{\,2}\left(  x\right)  $
is invariant under local infinitesimal gauge transformations. The momentum
current, $J_{P}\left(  x\right)  ,$ is self-invariant.

Once we have the invariant lagrangian we proceed for obtaining the
4quarks-$W^{\mu}$ vertex. The procedure is exactly the same as in the previous
appendix, expanding the currents in the number of $W$ mesons, and we write
directly our results.

The vertex associated to the interacting term $g_{0}\left(  J_{S}^{\left(
0W\right)  \dag}\left(  x\right)  J_{S}^{\left(  1W\right)  }\left(  x\right)
+J_{S}^{\left(  1W\right)  \dag}\left(  x\right)  J_{S}^{\left(  0W\right)
}\left(  x\right)  \right)  $ of the Lagrangian, in which a boson of type
$W^{1}$ is involved, is
\begin{gather}
\left[  \Gamma_{S,\mu}^{(4qW)}\left(  p_{1},p_{3};p_{2},p_{4}\right)  \right]
_{\alpha\beta,\gamma\delta}=-i~2g_{0}\frac{g_{w}}{2}G_{0}\left(  \frac
{p_{2}+p_{4}}{2}\right)  \delta_{\beta\delta}\left\{  \left[  \left(
p_{1}+p_{3}\right)  _{\mu}\mathbb{V}_{0a}\left(  \frac{p_{1}+p_{3}}%
{2},k\right)  \right.  \right. \nonumber\\
-\left.  k_{\mu}\mathbb{V}_{0b}\left(  \frac{p_{1}+p_{3}}{2},k\right)
\right]  \left(  \frac{1}{2}\tau^{1}\right)  _{\gamma\alpha}+\left[  \left(
p_{1}+p_{3}\right)  _{\mu}\mathbb{A}_{0a}\left(  \frac{p_{1}+p_{3}}%
{2},k\right)  \right. \nonumber\\
\left.  -\left.  k_{\mu}\mathbb{A}_{0b}\left(  \frac{p_{1}+p_{3}}{2},k\right)
\right]  \left(  \frac{1}{2}\tau^{1}\gamma_{5}\right)  _{\gamma\alpha
}\right\}  +\left(  p_{1}\alpha,p_{3}\gamma\leftrightarrow p_{2}\beta
,p_{4}\delta\right)  \ . \label{0B.15}%
\end{gather}
The vertex associated to the interacting term $g_{0}\left(  J_{5}^{\left(
0W\right)  \dag}\left(  x\right)  J_{5}^{\left(  1W\right)  }\left(  x\right)
+J_{5}^{\left(  1W\right)  \dag}\left(  x\right)  J_{5}^{\left(  0W\right)
}\left(  x\right)  \right)  $ of the Lagrangian is
\begin{gather}
\left[  \Gamma_{5,\mu}^{(4qW)}\left(  p_{1},p_{3};p_{2},p_{4}\right)  \right]
_{\alpha\beta,\gamma\delta}=-i~2g_{0}\frac{g_{w}}{2}G_{0}\left(  \frac
{p_{2}+p_{4}}{2}\right)  \left(  i\gamma_{5}\tau^{j}\right)  _{\beta\delta
}\left\{  i\left[  \left(  p_{1}+p_{3}\right)  _{\mu}\mathbb{V}_{0a}\left(
\frac{p_{1}+p_{3}}{2},k\right)  \right.  \right. \nonumber\\
-\left.  k_{\mu}\mathbb{V}_{0b}\left(  \frac{p_{1}+p_{3}}{2},k\right)
\right]  \left(  \frac{1}{2}\left\{  \tau^{1},\tau^{j}\right\}  \frac{1}%
{2}\gamma_{5}+\frac{1}{2}\left[  \tau^{1},\tau^{j}\right]  \frac{1}{2}\right)
_{\gamma\alpha}+\left[  \left(  p_{1}+p_{3}\right)  _{\mu}\mathbb{A}%
_{0a}\left(  \frac{p_{1}+p_{3}}{2},k\right)  \right. \nonumber\\
\left.  -\left.  k_{\mu}\mathbb{A}_{0b}\left(  \frac{p_{1}+p_{3}}{2},k\right)
\right]  \left(  \frac{1}{2}\left\{  \tau^{1},\tau^{j}\right\}  \frac{1}%
{2}+\frac{1}{2}\left[  \tau^{1},\tau^{j}\right]  \frac{1}{2}\gamma_{5}\right)
_{\gamma\alpha}\right\}  +\left(  p_{1}\alpha,p_{3}\gamma\leftrightarrow
p_{2}\beta,p_{4}\delta\right)  \ . \label{0B.16}%
\end{gather}
The vertex associated to the interacting term $g_{p}\left(  J_{p}^{\left(
0W\right)  \dag}\left(  x\right)  J_{p}^{\left(  1W\right)  }\left(  x\right)
+J_{p}^{\left(  1W\right)  \dag}\left(  x\right)  J_{p}^{\left(  0W\right)
}\left(  x\right)  \right)  $ of the Lagrangian is%
\begin{gather}
\left[  \Gamma_{p,\mu}^{(4qW)}\left(  p_{1},p_{3};p_{2},p_{4}\right)  \right]
_{\alpha\beta,\gamma\delta}=-i2g_{p}g_{w}G_{p}\left(  \frac{p_{2}+p_{4}}%
{2}\right)  \left(  \frac{\rlap{$/$}p_{2}+\rlap{$/$}p_{4}}{2}\right)
_{\delta\beta}\left(  \tau^{1}\gamma^{\nu}\frac{1}{2}\left(  1-\gamma
_{5}\right)  \right)  _{\gamma\alpha}\nonumber\\[0.1in]
\left\{  g_{\nu\mu}\frac{1}{2}\left[  G_{p}\left(  \frac{p_{1}+p_{3}+k}%
{2}\right)  +G_{p}\left(  \frac{p_{1}+p_{3}-k}{2}\right)  \right]  \right.
\nonumber\\
+\left.  \frac{\left(  p_{1}+p_{3}\right)  _{\nu}}{2}\left[  \left(
p_{1}+p_{3}\right)  _{\mu}\mathbb{V}_{pa}\left(  \frac{p_{1}+p_{3}}%
{2},k\right)  -k_{\mu}\mathbb{V}_{pb}\left(  \frac{p_{1}+p_{3}}{2},k\right)
\right]  \right\}  +\left(  p_{1}\alpha,p_{3}\gamma\leftrightarrow p_{2}%
\beta,p_{4}\delta\right)  \ , \label{0B.17}%
\end{gather}
In all these expressions, the momentum transferred is defined as
$k=p_{3}+p_{4}-p_{1}-p_{2}.$ The functions $\mathbb{V}_{0a}$, $\mathbb{V}%
_{0b}$, $\mathbb{A}_{0a}$, $\mathbb{A}_{0b}$, $\mathbb{V}_{pa}$ and
$\mathbb{V}_{pb}$ are defined in equations (\ref{0A.11a}), (\ref{0A.11b}),
(\ref{0A.16c}), (\ref{0A.16d}), (\ref{0A.21a}) and (\ref{0A.21b}).

The $W^{1}$ boson couples to quarks through the dressed vertex $\left(
\Gamma^{\mu}\left(  p_{1},p_{2}\right)  -\Gamma_{5}^{\mu}\left(  p_{1}%
,p_{2}\right)  \right)  \tau^{1}.$ As in the quark photon vertex discussed in
section \ref{SecQuarkPhotonVertex}, the full quark $W$ vertex is constructed
in a two steps process. The first one is shown in fig. \ref{FigQuarkPhoton0}
and it consists in the renormalization of the bare quark-$W$ vertex by the 4
quarks one $W$ vertex. The change in the vector part of the vertex is given in
equation (\ref{05.09DressedQuarkPhoton0}). The axial part is%
\begin{gather}
\Gamma_{5,0}^{\mu}\left(  p_{1},p_{2}\right)  =\gamma^{\mu}\gamma_{5}%
+\alpha_{0}\left[  \left(  p_{1}+p_{2}\right)  ^{\mu}\,\mathbb{A}_{0a}\left(
\bar{p},k\right)  +k^{\mu}\,\mathbb{A}_{0b}\left(  \bar{p},k\right)  \right]
\gamma_{5}+\nonumber\\
2~G_{0}\left(  \frac{p_{1}+p_{2}}{2}\right)  \frac{\left(  p_{1}-p_{2}\right)
^{\mu}}{\left(  p_{1}-p_{2}\right)  ^{2}}\gamma_{5}\left[  \alpha_{0}%
+g_{0}F_{1}\left(  k\right)  \right]  -\nonumber\\
\alpha_{p}\frac{1}{2}\left[  G_{p}\left(  p_{1}\right)  +G_{p}\left(
p_{2}\right)  \right]  \mathbb{\gamma}^{\mu}\gamma_{5}-\nonumber\\
\alpha_{p}\frac{\rlap{$/$}p_{1}+\rlap{$/$}p_{2}}{2}\gamma_{5}\left[  \left(
p_{1}+p_{2}\right)  ^{\mu}\,\mathbb{V}_{pa}\left(  \frac{p_{1}+p_{2}}%
{2},k\right)  +k^{\mu}\,\mathbb{V}_{pb}\left(  \frac{p_{1}+p_{2}}{2},k\right)
\right]  \label{0B.20}%
\end{gather}
with $F_{1}\left(  k\right)  $ given by equation (\ref{05a.23}). It is
interesting to note that the longitudinal part of $\Gamma_{5,0}^{\mu}\left(
p_{1},p_{2}\right)  $,%
\begin{equation}
\left(  p_{2}-p_{1}\right)  _{\mu}\Gamma_{5,0}^{\mu}\left(  p_{1}%
,p_{2}\right)  =S^{-1}\left(  p_{2}\right)  \gamma_{5}+\gamma_{5}S^{-1}\left(
p_{1}\right)  +\left(  2m_{0}-2g_{0}G_{0}\left(  \frac{p_{1}+p_{2}}{2}\right)
F_{1}\left(  k^{2}\right)  \right)  \gamma_{5}~, \label{0B.21}%
\end{equation}
does not satisfy the WTI (\ref{05a.01}).

The second step is represented in Fig. \ref{FigQuarkPhoton1}. The associated
equation for the vector part of the vertex is given in equation
(\ref{05.10DSE-QuarkPhoton}) and the final form of the solution is given in
equation (\ref{05.12DressedQuarkPhoton}). The axial part of the vertex is
governed by the equation%
\begin{gather}
i\,\Gamma_{5}^{\mu}\left(  p_{1},p_{2}\right)  \frac{\tau^{1}}{2}%
=i\,\Gamma_{5,0}^{\mu}\left(  p_{1},p_{2}\right)  \frac{\tau^{1}}%
{2}+\nonumber\\
i\,2\,g_{0}\,G_{0}\left(  \frac{p_{1}+p_{2}}{2}\right)  \,i\gamma_{5}\tau
^{j}\,\int\frac{d^{4}p}{\left(  2\pi\right)  ^{4}}G_{0}\left(  p\right)
\,\left(  -\right)  \mathbb{T}\text{r}\left[  i\,S\left(  p-\frac{k}%
{2}\right)  \,i\gamma_{5}\,\tau^{j}\,i\,S\left(  p+\frac{k}{2}\right)
\,i\,\Gamma_{5}^{\mu}\left(  p-\frac{k}{2},p+\frac{k}{2}\right)  \frac
{\tau^{1}}{2}\right]  \label{0B.22}%
\end{gather}
with $k=p_{2}-p_{1}.$ This equation can be solved obtaining%
\begin{equation}
\Gamma_{5}^{\mu}\left(  p_{1},p_{2}\right)  =\Gamma_{5,0}^{\mu}\left(
p_{1},p_{2}\right)  +2g_{0}G_{0}\left(  \frac{p_{1}+p_{2}}{2}\right)  \left[
F_{1}\left(  k^{2}\right)  +2m_{0}\frac{F_{0}\left(  k^{2}\right)  }%
{1-2g_{0}~\Pi_{PS}\left(  k^{2}\right)  }\right]  \frac{k^{\mu}}{k^{2}}%
\gamma_{5} \label{0B.25}%
\end{equation}
This expression can be rewritten in the form given by equation (\ref{05a.10}).

\bigskip

In equation (\ref{05a.31}) we give an approximated expression for the
longitudinal part of $\Gamma_{5,0}^{\mu}\left(  p_{1},p_{2}\right)  .$ From
equations (\ref{05a.31}) and (\ref{0B.21}) we have%
\begin{gather}
\left(  p_{1}-p_{2}\right)  _{\mu}\left(  \Gamma_{5,0}^{\mu}\left(
p_{1},p_{2}\right)  -\Gamma_{5,0}^{\prime\mu}\left(  p_{1},p_{2}\right)
\right)  =\nonumber\\
\left[  \alpha_{0}~G_{0}\left(  p_{1}\right)  +\alpha_{0}~G_{0}\left(
p_{2}\right)  +2g_{0}~G_{0}\left(  p\right)  ~F_{1}\left(  P^{2}\right)
\right]  \gamma_{5} \label{0B.26}%
\end{gather}
with $p_{1,2}=p\pm P/2.$ Inserting equation (\ref{0B.26}) in (\ref{05a.13}) we
can evaluate the numerical error produced in the calculation of $f_{\pi}$
through the use of $\Gamma_{5,0}^{\prime\mu}\left(  p_{1},p_{2}\right)  .$
Expanding in powers of $P^{2}$ it is straightforward to obtain that $\Delta
f_{\pi}=\mathcal{O}\left(  m_{\pi}^{2}\right)  .$

There are alternatives ways for calculating the pion decay constant in which
we cannot use the approximated expression (\ref{05a.31}). For instance, we can
consider an interacting pair $q\bar{q}$ which a some point couples to a $W$
boson. We can describe the interaction between the $q\bar{q}$ by the
scattering amplitude or using the dressed vertex. This two descriptions must
be equivalents and in the proximity of the pion pole we have%
\begin{gather}
iP^{\mu}f_{\pi}\frac{i}{P^{2}-m_{\pi}^{2}}\Gamma_{\gamma\alpha}^{\pi}\left(
p^{\prime},P\right)  =\nonumber\\
\left(  -\right)  \int\frac{d^{4}p}{\left(  2\pi\right)  ^{4}}2iG_{\alpha
\beta\gamma\delta}\left(  p^{\prime},p,P\right)  \left(  i\,S\left(
p-\frac{1}{2}P\right)  \,\Gamma_{5}^{\mu}\left(  p+\frac{1}{2}P,p-\frac{1}%
{2}P\right)  \,\frac{\tau^{j}}{2}\,i\,S\left(  p+\frac{1}{2}P\right)  \right)
_{\beta\delta} \label{0B.30}%
\end{gather}
Where, in our model, the interaction is%
\begin{equation}
G_{\alpha\beta\gamma\delta}\left(  p^{\prime},p,P\right)  =g_{0}\ G_{0}\left(
p^{\prime}\right)  G_{0}\left(  p\right)  \left[  \delta_{\gamma\alpha}%
\delta_{\delta\beta}+\left(  i\vec{\tau}\gamma_{5}\right)  _{\gamma\alpha
}\left(  i\vec{\tau}\gamma_{5}\right)  _{\delta\beta}\right]  +g_{p}%
\ G_{p}\left(  p^{\prime}\right)  G_{p}\left(  p\right)  \left(
\rlap{$/$}p^{\prime}\right)  _{\gamma\alpha}\left(  \rlap{$/$}p\right)
_{\delta\beta}~~, \label{0B.31}%
\end{equation}
and the pion amplitude is
\begin{equation}
\Gamma_{\alpha\gamma}^{\pi}\left(  p,P\right)  =i\,g_{\pi qq}\,G_{0}\left(
p\right)  \,\left(  i\,\gamma_{5}\,\tau^{i}\right)  _{\alpha\gamma}
\label{0B.32}%
\end{equation}
In the right hand side of equation (\ref{0B.30}) we must consider only the
pion pole contribution. It is easy to reproduced the result obtained in
equation (\ref{05a.22}) for $f_{\pi}$, and is also obvious that we cannot use
of the expression (\ref{05a.31}), because we lost the pion pole in the right
hand side of equation (\ref{0B.30}).

\section{Four quarks-photon vertex in the general case and the value of the
form factor at $k^{2}=0$.}

\label{App4q-fGeneralVetrex}

The standard normalization condition for the Bethe-Salpeter amplitude is
\begin{gather}
\hspace*{-13cm}2iP^{\mu}=\nonumber\\
\int\frac{d^{4}p}{\left(  2\pi\right)  ^{4}}\mathbb{T}\text{r}\left[
\bar{\Gamma}^{M}\left(  p,P\right)  i\,\frac{\partial S\left(  p+\frac{1}%
{2}P\right)  }{\partial P_{\mu}}\,\Gamma^{M}\left(  p,P\right)  \,i\,S\left(
p-\frac{1}{2}P\right)  +\bar{\Gamma}^{M}\left(  p,P\right)  i\,S\left(
p+\frac{1}{2}P\right)  \,\Gamma^{M}\left(  p,P\right)  \,i\,\frac{\partial
S\left(  p-\frac{1}{2}P\right)  }{\partial P_{\mu}}\right] \nonumber\\
-2i\int\frac{d^{4}p}{\left(  2\pi\right)  ^{4}}\frac{d^{4}p^{\prime}}{\left(
2\pi\right)  ^{4}}\left[  i\,S\left(  p-\frac{1}{2}P\right)  \bar{\Gamma}%
^{M}\left(  p,P\right)  i\,S\left(  p+\frac{1}{2}P\right)  \right]
_{\alpha\gamma}\frac{\partial G_{\alpha\beta\gamma\delta}\left(  p,p^{\prime
},P\right)  }{\partial P_{\mu}}\nonumber\\
\left[  i\,S\left(  p^{\prime}+\frac{1}{2}P\right)  \,\Gamma^{M}\left(
p^{\prime},P\right)  \,i\,S\left(  p^{\prime}-\frac{1}{2}P\right)  \right]
_{\beta\delta}\ \ , \label{07.06.normalizacion}%
\end{gather}
where
\begin{equation}
\,\bar{\Gamma}^{M}\left(  p,P\right)  =\gamma_{0}\left[  \Gamma^{M}\left(
p,P\right)  \right]  ^{\dag}\gamma_{0}\ .
\end{equation}
This normalization condition is equivalent to
equation(\ref{04.06NormaBoundState}).

A minimal test of consistency of our calculation is that the form factor at
$k^{2}=0$ must be 1 for an amplitude normalized with equation
(\ref{07.06.normalizacion}). Usually the form factor is calculated in the
impulse approximation, which includes only the triangle diagram shown in
Fig.~\ref{FigPionFF1}. The use of the Ward identity (\ref{07.00}) in order to
define the electromagnetic vertex in this diagram provides a contribution
which coincides with the first integral on the right hand side of equation
(\ref{07.06.normalizacion}). From that we can conclude that the use of the BSE
for the pion with a $P$-independent Bethe-Salpeter kernel together with the
triangle diagram provides a consistent approximation scheme \cite{Roberts96}.
For a $P$-dependent kernel this consistency is lost even at $k^{2}=0$ due to
the presence of the second integral on the right hand side of equation
(\ref{07.06.normalizacion}).

Let us proof that $F\left(  0\right)  =1$ from equation (\ref{07.08}). We need
to expand equation (\ref{07.08}) in powers of the photon field. We retain the
second term, which is linear in the photon field. We can evaluate it in the
limit $k^{2}\rightarrow0$ without defining a specific path for the integrals
present in equation(\ref{07.08}), obtaining
\begin{gather}
\left[  Q_{1}\int_{X}^{\frac{x^{\prime}-X^{\prime}}{2}+X}dz^{\mu}A_{\mu
}\left(  z\right)  +Q_{2}\int_{-\frac{x^{\prime}+X^{\prime}}{2}+X}^{X}dz^{\mu
}A_{\mu}\left(  z\right)  +Q_{2}\int_{X}^{\frac{x+X^{\prime}}{2}+X}dz^{\mu
}A_{\mu}\left(  z\right)  +Q_{1}\int_{-\frac{x-X^{\prime}}{2}+X}^{X}dz^{\mu
}A_{\mu}\left(  z\right)  \right] \nonumber\\
\underset{k^{2}\rightarrow0}{\longrightarrow}\epsilon_{\mu}\left(
k,\xi\right)  \left[  -\left(  Q_{1}-Q_{2}\right)  X^{\prime\mu}+\frac
{Q_{1}+Q_{2}}{2}\left(  x^{\mu}+x^{\prime\mu}\right)  \right]  \ .
\end{gather}
From this last result it is easy to see that the quantity to add to each
vertex of the type of Fig.~\ref{Fig4quarkphoton} in the limit of
$k^{2}\rightarrow0$, is given by Eqs. (\ref{07.11}) and (\ref{07.12}).

The electromagnetic form factor in the limit $k^{2}\rightarrow0,$ including
the contributions from Figs.\ref{FigPionFF1} and \ref{FigPionFF2}, is%

\begin{gather}
i\,e\,2\,\bar{P}^{\mu}F\left(  0\right) \nonumber\\[0.2cm]
=-iQ_{2}\int\frac{d^{4}p}{\left(  2\pi\right)  ^{4}}\,\mathbb{T}%
\text{r}\left[  \Gamma^{M}\left(  p,P\right)  i\,S\left(  p-\frac{1}%
{2}P\right)  \right.  \left.  \,\,\bar{\Gamma}^{M}\left(  p,P\right)
\,i\,S\left(  p+\frac{1}{2}P\right)  \,\,\Gamma^{\mu}\left(  p+\frac{1}%
{2}P,p+\frac{1}{2}P\right)  \,i\,S\left(  p+\frac{1}{2}P\right)  \right]
\nonumber\\
-\,i\,Q_{1}\int\frac{d^{4}p}{\left(  2\pi\right)  ^{4}}\,\mathbb{T}%
\text{r}\left[  \Gamma^{M}\left(  p,P\right)  \,i\,S\left(  p-\frac{1}%
{2}P\right)  \,\Gamma^{\mu}\left(  p-\frac{1}{2}P,p-\frac{1}{2}P\right)
\right.  \left.  i\,S\left(  p-\frac{1}{2}P\right)  \,\bar{\Gamma}^{M}\left(
p,P\right)  \,i\,S\left(  p+\frac{1}{2}P\right)  \right] \nonumber\\
-\int\frac{d^{4}p}{\left(  2\pi\right)  ^{4}}\int\frac{d^{4}p^{\prime}%
}{\left(  2\pi\right)  ^{4}}\,\left[  i\,S\left(  p^{\prime}-\frac{1}%
{2}P\right)  \bar{\Gamma}^{M}\left(  p^{\prime},P\right)  i\,S\left(
p^{\prime}+\frac{1}{2}P\right)  \right]  _{\alpha\gamma}\nonumber\\
\left\{  -iQ_{1}\left[  \Gamma_{1,\mu}^{(4q\gamma)}\left(  p^{\prime}-\frac
{1}{2}P,p^{\prime}+\frac{1}{2}P;p+\frac{1}{2}P,p-\frac{1}{2}P\right)  \right]
_{\alpha\beta,\gamma\delta}\right. \nonumber\\
\left.  -iQ_{2}\left[  \Gamma_{2,\mu}^{(4q\gamma)}\left(  p^{\prime}-\frac
{1}{2}P,p^{\prime}+\frac{1}{2}P;p+\frac{1}{2}P,p-\frac{1}{2}P\right)  \right]
_{\alpha\beta,\gamma\delta}\right\}  \left[  \,i\,S\left(  p+\frac{1}%
{2}P\right)  \Gamma^{M}\left(  p,P\right)  i\,S\left(  p-\frac{1}{2}P\right)
\right]  _{\beta\delta}\ . \label{07.14}%
\end{gather}
In order to simplify the first two lines of this equation (the one body part)
we make use of the WTI, equation (\ref{07.00}). For the two body part of the
equation we use
\begin{align}
\left[  -iQ_{1}\Gamma_{1,\mu}^{(4q\gamma)}\left(  p^{\prime}-\frac{1}%
{2}P,p^{\prime}+\frac{1}{2}P;p+\frac{1}{2}P,p-\frac{1}{2}P\right)  \right]
_{\alpha\beta,\gamma\delta}  &  =-iQ_{1}\left(  \frac{d}{dp^{\prime\mu}}%
+\frac{d}{dp^{\mu}}-2\frac{d}{dP^{\mu}}\right)  G_{\alpha\beta\gamma\delta
}\left(  p^{\prime},p,P\right)  \ ,\label{07.14a}\\[0.3cm]
\left[  -iQ_{2}\Gamma_{2,\mu}^{(4q\gamma)}\left(  p^{\prime}-\frac{1}%
{2}P,p^{\prime}+\frac{1}{2}P;p+\frac{1}{2}P,p-\frac{1}{2}P\right)  \right]
_{\alpha\beta,\gamma\delta}  &  =-iQ_{2}\left(  \frac{d}{dp^{\prime\mu}}%
+\frac{d}{dp^{\mu}}+2\frac{d}{dP^{\mu}}\right)  G_{\alpha\beta\gamma\delta
}\left(  p^{\prime},p,P\right)  \ . \label{07.14b}%
\end{align}
Equation (\ref{07.05BSE}) allows to do some of these integrals. Charge
conjugation symmetry leads to
\begin{align}
&  \int\frac{d^{4}p}{\left(  2\pi\right)  ^{4}}\mathbb{T}\text{r}\left[
\frac{d~\bar{\Gamma}^{M}\left(  p,P\right)  }{dp^{\mu}}\,i\,S\left(
p+\frac{1}{2}P\right)  \Gamma^{M}\left(  p,P\right)  i\,S\left(  p-\frac{1}%
{2}P\right)  \right] \nonumber\\
&  =\int\frac{d^{4}p}{\left(  2\pi\right)  ^{4}}\mathbb{T}\text{r}\left[
i\,S\left(  p-\frac{1}{2}P\right)  \bar{\Gamma}^{M}\left(  p,P\right)
i\,S\left(  p+\frac{1}{2}P\right)  \frac{d~\Gamma^{M}\left(  p,P\right)
}{dp^{\mu}}\right]  =0\ \ . \label{07.15}%
\end{align}
With these inputs, the normalization condition of the Bethe-Salpeter
amplitude, equation (\ref{07.06.normalizacion}), implies the natural
normalization for the form factor, $F\left(  0\right)  =1.$ We observe that
only the contribution associated with the derivative of the total momentum in
Eqs.(\ref{07.14a}) and (\ref{07.14b}) gives a non vanishing result.

Our results show that consistency between the Bethe-Salpeter normalization
condition equation (\ref{07.06.normalizacion}) and the value of the meson form
factor at $k^{2}=0$ is also attainable for a $P$-dependent kernel, if we add
the contributions coming from the diagrams of Figs.~\ref{FigPionFF1} and
\ref{FigPionFF2}. This result is consistent with field theory. We have simply
added all the diagrams with a one photon coupling, no matter where the photon
couples in our system, and in this way we have obtained the gauge invariant
contribution to the form factor. Fig.~\ref{FigPionFF2} confirms that the use
of the Ward-Takahashi identities for the components of a system is not
sufficient to assure that the gauge symmetry is satisfied for the composite system.

\end{document}